\documentclass[12pt]{article}
\usepackage{amsmath,amssymb,bbold,epsfig}
\usepackage{amsfonts}
\usepackage{bbm}
\renewcommand{\vec}[1]{{\mathbf{#1}}}

\newcommand{\be}{\begin{equation}}
\newcommand{\ee}{\end{equation}}
\newcommand{\bea}{\begin{eqnarray}}
\newcommand{\eea}{\end{eqnarray}}
\begin{document}

\title{
\Large \bf
Neutrino production coherence and oscillation experiments}
\author{
{E. Kh. Akhmedov$^{a,b}$\thanks{email: \tt 
akhmedov@mpi-hd.mpg.de},~~\,D. Hernandez$^{c}$\thanks{email:
\tt dhernand@ictp.it}~~\,and
\vspace*{0.15cm} ~A. Yu. Smirnov
$^{c}$\thanks{email:
\tt smirnov@ictp.it}
} \\
{\normalsize\em $^a$Max--Planck--Institut f\"ur Kernphysik,
Postfach 103980} \\ {\normalsize\em D--69029 Heidelberg, Germany
\vspace*{0.15cm}}
\\
{\normalsize\em $^{b}$National Research Centre Kurchatov
\vspace*{-0.1cm}Institute}\\{\normalsize\em Moscow, Russia 
\vspace*{0.15cm}}
\\
{\normalsize\em $^{c}$The Abdus Salam International Centre for Theoretical    
Physics} \\
{\normalsize\em Strada Costiera 11, I-34014 Trieste, Italy 
\vspace*{0.15cm}}
}
\date{}
\maketitle
\thispagestyle{empty}
\vspace{-0.8cm}
\begin{abstract}
Neutrino oscillations are only observable when the neutrino production, 
propagation and detection coherence conditions are satisfied. In this paper 
we consider in detail neutrino production coherence, taking $\pi\to \mu \nu$ 
decay as an example. We compare the oscillation probabilities obtained in two 
different ways: (1) coherent summation of the amplitudes of neutrino 
production at different points along the trajectory of the parent pion; 
(2) averaging of the standard oscillation probability over the neutrino 
production coordinate in the source. We demonstrate that the results of 
these two different approaches exactly coincide, provided that the parent 
pion is considered as pointlike and the detection process is perfectly 
localized. In this case the standard averaging of the oscillation probability 
over the finite spatial extensions of the neutrino source (and detector) 
properly takes possible decoherence effects into account. We analyze 
the reason for this equivalence of the two approaches and demonstrate that 
for pion wave packets of finite width $\sigma_{x\pi}$ the equivalence is 
broken. The leading order correction to the oscillation probability due 
to $\sigma_{x\pi}\ne 0$ is shown to be $\sim [v_g/(v_g-v_\pi)]\sigma_{x\pi}
/l_{\rm osc}$, where $v_g$ and $v_\pi$ are the group velocities of the 
neutrino and pion wave packets, and $l_{\rm osc}$ is the neutrino 
oscillation length.
 
\end{abstract}

\vspace{1.cm}
\vspace{.3cm}

\newpage

\tableofcontents

\section{\label{sec:intro}Introduction}

It is well known that neutrino oscillations are only observable when 
the conditions of coherent production, propagation and detection of 
different neutrino mass eigenstates are satisfied. The production and 
detection coherence conditions 
ensure that the intrinsic quantum-mechanical energy uncertainties at 
neutrino production and detection are large compared to the energy 
difference $\Delta E_{jk}$ of different neutrino mass eigenstates:
\be
\Delta E_{jk}\sim \frac{\Delta m_{jk}^2}{2E}\ll \sigma_E\,, 
\label{eq:coh1}
\ee
where $\sigma_E=\min\{\sigma_E^{\rm prod},\,\sigma_E^{\rm det}\}$. 
If this condition is violated, the neutrino production or detection process 
will be able to discriminate between different neutrino mass eigenstates, 
thus destroying neutrino oscillations. Indeed, neutrino mass eigenstates do 
not oscillate in vacuum, and the oscillations are only possible because 
neutrinos are emitted and detected as flavour states -- coherent linear 
superpositions of different mass eigenstates. This coherence is destroyed 
if the condition (\ref{eq:coh1}) is violated. Production/detection coherence 
is known to be related to the localization of the corresponding neutrino 
emission and absorption processes, as it is the localization of these 
processes that determines the energy uncertainty $\sigma_E$ 
\cite{Kayser}.

The propagation decoherence can take place if neutrinos propagate very long 
distances. It is related to the fact that the wave packets describing the  
different neutrino mass eigenstates that compose a flavour state propagate 
with different group velocities, and therefore after a long enough time they 
cease to overlap and separate to such an extent that 
the amplitudes of their interaction with the detector particles cannot 
interfere. For ultrarelativistic neutrinos the propagation coherence 
condition can be written as    
\be
\frac{\Delta m_{jk}^2}{2E^2}L \ll \sigma_{x\nu}\simeq v_g/\sigma_E\,,
\label{eq:coh2}
\ee
where $L$ is the baseline, $v_g$ is the average group velocity of the wave 
packets of different neutrino mass eigenstates and $\sigma_{x\nu}$
is their common effective spatial width.
Note that both the production/detection and propagation coherence 
conditions, eqs.~(\ref{eq:coh1}) and~(\ref{eq:coh2}), put upper limits on 
the mass squared differences $\Delta m_{jk}^2$ of different neutrino mass 
eigenstates.

In practice, the propagation coherence condition is very well satisfied 
in all cases except for neutrinos of astrophysical or cosmological origin, 
such as solar, supernova or relic neutrinos. As to the production/detection 
coherence condition, it is usually tacitly assumed to be always satisfied 
due to the extreme smallness of the neutrino masses (and therefore of their 
mass squared differences). However, there may exist 
situations in which this is not the case. 
In particular, in view of possible existence of sterile neutrinos with 
masses in the eV, keV or even MeV range, the question of
production/detection coherence should be analyzed especially carefully
when the corresponding mass squared differences are involved.
Note that there are some hints for such sterile neutrinos coming from
short-baseline accelerator experiments, the reactor neutrino anomaly, 
gallium radiative source experiments, $r$-process supernova
nucleosynthesis, pulsar kicks, warm dark matter and leptogenesis 
scenarios \cite{SNAC11}.

The issue of coherence is of utmost importance for neutrino oscillation 
experiments. If the coherence conditions are strongly violated, 
the probabilities of flavour transitions will correspond to averaged out  
oscillations, i.e.\ they will have neither $L$ nor $E$ 
dependence. This means that the two most important signatures of neutrino 
oscillations -- the dependence on the distance between the neutrino source 
and detector and the distortion of the energy spectrum of the detected 
signal -- will be absent. In particular, the two-detector setups of  
neutrino oscillation experiments will be completely useless.  In such 
situations one would have to rely entirely on the overall normalization 
of the neutrino flux, which is usually known with insufficient accuracy. 

The question of neutrino production coherence has been recently addressed  
in ref.~\cite{HS}, where it was considered in the framework of incoherent 
summation of probabilities of neutrino production at different points inside 
the neutrino source. In addition, a simplified study of the coherent amplitude 
summation was performed in that paper. It was demonstrated that under 
certain conditions the two summation procedures lead to identical results.   

In this paper we address this question within a more rigorous and 
consistent approach. We study in detail neutrino production coherence, 
taking $\pi\to \mu \nu$ decay as an example. With minimal modifications, 
our analysis will also be applicable to neutrino detection. We find the 
oscillations probabilities in two different approaches: 
\begin{itemize}
\item[1.]
An approach based on the quantum-mechanical wave packet formalism. We 
first calculate the transition amplitude by summing the amplitudes 
corresponding to neutrino production in pion decay at different points 
along the trajectory of the parent pion. We then calculate the transition 
probability and study its coherence properties. 
\item[2.]
We assume that for each individual neutrino 
production event the oscillation probability is fully coherent but depends 
on the exact position of the neutrino production point, i.e.\ is 
described by the standard oscillation formula. The effective oscillation 
probability is then found by integrating (averaging) this standard 
coordinate-dependent probability along the neutrino source, with the proper 
exponential factor describing the pion decay included \cite{HS}. 
\end{itemize}
{\sl
It should be stressed that, strictly speaking, the first approach, based 
on a quantum mechanical amplitude summation, should always be used. 
The second (probability) summation procedure, which is classical in 
nature, is however 
much simpler and is usually employed in the analyses of experiments. 
It is therefore one of the main goals of the present paper to study if 
and when the use of the probability summation approach is indeed justified.}

We compare the results of the two approaches and demonstrate that they exactly 
coincide when the parent pions are considered as pointlike particles and the 
detection process is perfectly localized in space and time. One can therefore 
conclude that for pointlike parent particles and well-localized detection the 
standard averaging of the oscillation probabilities over the finite spatial 
extensions of the neutrino source (and detector) properly takes possible 
decoherence effects into account. We analyze the reason for this equivalence 
of the two approaches and demonstrate that for finite-size pion wave packets 
the equivalence is broken. 
We show that for small widths of the pion wave packets $\sigma_{x\pi}$ the 
oscillation probabilities get small oscillatory corrections which are linear 
in $\sigma_{x\pi}$. For different shapes of the pion wave packets these 
corrections have the same form and differ only by numerical coefficients. 
At the same time, large $\sigma_{x\pi}$ can lead to production decoherence and 
thus to suppression of the oscillations. We also consider the production 
coherence in the case when the charged leptons accompanying neutrino 
production are detected, leading to neutrino tagging, as well as in the case 
when the interactions of pions in the bunch between themselves or with other 
particles which may be present in the neutrino source are taken into account. 

{\bf A note on terminology.} In this paper we use the word `coherence' 
in two different, though related, senses. By {\em production coherence} we 
mean that different neutrino {\em mass eigenstates} are produced coherently, 
so that the emitted neutrino is a flavour eigenstate. By {\em coherent 
summation} we mean the summation of the amplitudes of neutrino production at 
{\em different points} along the trajectory of the parent pion. As we 
shall see, these two coherences are in fact acting in opposite 
directions: if the production is incoherent, then the coherent summation 
(i.e.\ approach 1 discussed above) is mandatory, whereas for coherent neutrino 
production the incoherent probability summation (approach 2) is justified. 
The common feature of the two coherences 
is that both require summation of certain amplitudes. Indeed, in the case 
of neutrino production coherence, these  are the amplitudes of emission 
of different neutrino mass eigenstates, whereas in the case of 
coherent summation approach, these are the amplitudes of neutrino 
emission from different space-time points.  

The paper is organized as follows. In section \ref{sec:cohsum} we calculate 
the oscillation probability for neutrinos produced in decays of pointlike pions 
in a decay tunnel. The calculations are performed in the quantum-mechanical 
wave packet approach with summation of the amplitudes of neutrino production 
at different points along the tunnel. 
In \mbox{section \ref{sec:incoher}} we recapitulate how the same problem is 
solved at the level of summation of probabilities rather than amplitudes. In 
section \ref{sec:finlength} we generalize the results of the amplitude 
summation approach to the case of non-zero spatial width of the pion wave 
packets. We then consider the particular cases of Gaussian and box-type pion 
wave packets. In section \ref{sec:thick} we apply the probability  
summation approach to the case of protons incident on a finite-thickness 
target. We also compare the obtained results with those of section 
\ref{sec:finlength}. In section \ref{sec:altern} we develop an alternative 
approach to the quantum mechanical amplitude summation calculation, 
which is valid for arbitrary shapes of the pion wave packets. 
In \mbox{section \ref{sec:mupi}} we consider effects 
of possible detection of the charged lepton accompanying the neutrino 
production on neutrino production coherence. We also briefly discuss here 
the effects of the interaction of pions in the bunch between themselves or 
with other particles which may be present in the neutrino source.  
Section \ref{sec:exp} is devoted to implications of our analysis for 
various  neutrino experiments. We summarize and discuss our results in 
\mbox{section \ref{sec:disc}}.

\section{\label{sec:cohsum} Oscillation probabilities in the wave packet 
approach. Pointlike parent particle approximation}

In the quantum-mechanical wave packet approach, the oscillation amplitude 
${\cal A}_{\alpha\beta}(L,t)$ is obtained by projecting the evolved neutrino 
state, which was initially produced as the flavour eigenstate $\nu_\alpha$, 
onto the detected neutrino flavour eigenstate $\nu_\beta$. The oscillation 
probability $P_{\alpha\beta}(L)$ is then found by integrating the squared 
modulus of ${\cal A}_{\alpha\beta}(L,t)$ over time (see, e.g., \cite{AS1}). 

Let us first consider oscillations of neutrinos produced in decays of 
quasi-free parent particles. By this we mean that the decaying 
particles may be confined to a finite-size source or a decay tunnel, but 
their interactions with each other or with other particles which may be 
present within the source can be neglected (we will relax these assumptions 
in section \ref{sec:mupi}). As an example, we take the $\pi \to \mu\nu$ decay.%
\footnote{A similar (though somewhat different and less detailed) analysis 
was performed in \cite{Rich}.}
We will treat this problem in a 1-dimensional approach, i.e.\ assuming that 
the neutrino momentum $\vec{p}$ is parallel to the baseline vector $\vec{L}$
connecting the pion production and neutrino detection points.
Such an approximation is well justified when the transverse (i.e.\ 
orthogonal to $\vec{L}$) sizes of the neutrino source and detector are 
small compared to $L$.

To find the wave function of the produced neutrino state, we will need the 
wave functions of the parent pion and the muon which 
participate in the production process. 

\subsection{\label{sec:pimu}The pion and muon wave functions}

We will be assuming that pions are produced by a beam of protons incident 
on a solid-state target and then decay inside a decay tunnel, with the 
total length of the decay tunnel being $l_p$. 
Such a setup corresponds to accelerator experiments. The nuclei in the 
target are well localized, with the uncertainty of their position 
$\sigma_{xN}$ being of the order of inter-atomic distances, $\sigma_{xN}\sim 
10^{-8}$ cm. The spatial width of the wave packets of the incident protons 
$\sigma_{xp}$ depends on the conditions of their production; it cannot 
exceed the mean distance between the protons in the bunch, and e.g.\ for the 
Fermilab NuMI source it can be estimated as $\sigma_{xp}\lesssim 10^{-4}$ cm. 
The spatial size of the pion production region $\sigma_{xP}$ can be defined as 
\cite{Giunti1,Beuthe}
\be
\frac{1}{\sigma_{xP}^2}\equiv \frac{1}{\sigma_{xN}^2}
+\frac{1}{\sigma_{xp}^2}\,.
\label{eq:rel01}
\ee
We can also define the effective 
velocity of the pion production region as the sum of the group velocities $v_p$ 
and $v_N$ of the incident proton and the target nucleus, weighted with the 
inverse squared widths of the corresponding wave packets:  
\be
v_P\equiv \sigma_{xP}^2\left(\frac{v_p}{\sigma_{xp}^2}
+\frac{v_N}{\sigma_{xN}^2}\right).
\label{eq:rel02}
\ee
The coordinate-space width of the pion wave packets can then be found 
from the relation \cite{Giunti1,Beuthe,AK}
\be
\sigma_{x\pi}\approx [\sigma_{xP}^2+(v_P-v_\pi)^2/\sigma_{e\pi}^2]^{1/2}\,,
\label{eq:rel0}
\ee
where $v_\pi$ and $\sigma_{e\pi}$ are, respectively, the group velocity of the 
pion wave packet and the energy uncertainty of the produced pion state. The 
latter is approximately equal to the inverse of the overlap time of the proton 
and nucleon wave packets at pion production \cite{Beuthe}: 
\be
\sigma_{e\pi} \approx 
\frac{v_p-v_N}{(\sigma_{xp}^2+\sigma_{xN}^2)^{1/2}}        
\approx \frac{v_p}{\sigma_{xp}}\,.
\label{eq:epi}
\ee
Eq.~(\ref{eq:rel0}) thus has a simple physical meaning: the first term 
in the square brackets is the contribution of the finite spatial size of 
the pion production region to the width of the pion wave packet, 
whereas the second term is related to the fact that the pion production 
takes finite time.  

As follows from (\ref{eq:rel01}), the coordinate uncertainty of the pion 
production point $\sigma_{xP}$ is dominated by the size of the shortest wave 
packet of the participating particles, which in our case is $\sigma_{xN}$: 
$\sigma_{xP}\simeq \sigma_{xN}$. 
We shall also assume that the target nuclei are at rest, $v_N=0$. 
Eq.~(\ref{eq:rel02}) then gives $v_P\approx v_p(\sigma_{xN}/\sigma_{xp})^2 
\ll v_p$. From eqs.~(\ref{eq:rel0}) and (\ref{eq:epi}) we find 
\be              
\sigma_{x\pi} \approx \left[\sigma_{xN}^2+(v_\pi/v_p)^2\,\sigma_{xp}^2
\right]^{1/2}\,.
\label{eq:sigmaxpi}
\ee
Thus, $\sigma_{x\pi}\lesssim \sigma_{xp}\lesssim 10^{-4}$ cm, i.e.\ the 
spatial width of the pion wave packet is much 
smaller than all the lengths of interest in the problem -- the length of 
the decay tunnel $l_p$, the baseline $L$ and the oscillation length 
$l_{\rm osc}=4\pi p/\Delta m^2$, where $p$ is the neutrino momentum.  
Therefore to a very good approximation one can consider the pions as 
pointlike particles. 

Let us recall that the coordinate-space wave packet describing a 
moving free particle can be written, in the approximation where the 
spreading of the wave packet is neglected, as  
\be
\psi(x,t)\simeq e^{i p_0 x - iE(p_0)t} g(x-v_g t)\,,
\label{eq:wpgen}
\ee
where $p_0$ is the peak momentum of the wave packet, $v_g=(\partial 
E(p)/\partial p)|_{p_0}$ is its group velocity and $g(x-v_g t)$ is its 
shape factor (envelope function) (see, e.g., \cite{Bohm}). The shape 
factor is the Fourier transform of the momentum distribution amplitude 
(i.e.\ of the momentum-space wave function) of the pion; it quickly 
decreases when $|x-v_g t|$ becomes large compared to the spatial width of 
the wave packet $\sigma_x$. Since $g$ depends on $x$ and $t$ only through 
the combination $x-v_g t$, for stable particles eq.~(\ref{eq:wpgen}) 
describes a wave packet propagating with the group velocity $v_g$ without 
changing its shape. For unstable particles the energy $E(p_0)$ in 
eq.~(\ref{eq:wpgen}) should be replaced according to $E(p_0)\to 
E(p_0)-i\Gamma/2$, where $\Gamma$ is the particle's decay width in the 
laboratory frame.  

As was pointed out above, in the problem under consideration pions can 
to a very good approximation be regarded as pointlike particles, i.e. 
the shape factor $g_\pi(x-v_\pi t)$ of their wave packets can be 
taken to be a $\delta$-function. 
The pion 
wave function then takes the form 
\be
\psi_\pi(x,t)=C_\pi\,e^{iQx-iE_\pi(Q)t-\Gamma
t/2}\,\delta(x-v_\pi t)\,{\rm box}(x; l_p, 0)\,,
\label{eq:wppi}
\ee
where $C_\pi$ is a normalization constant, $Q$ is the mean momentum 
of the pion state,  $E_\pi(Q)=(Q^2+m_\pi^2)^{1/2}$, $v_\pi$ is the group 
velocity of the pion wave packet, and the function ${\rm box}(x; A, B)$ is 
defined as
\be
{\rm box}(x; A, B)=\left\{\begin{array}{ll}
1\,,\quad A\ge x \ge B\,, \\
0\,,\quad{\rm otherwise}
\end{array}
\right. .
\label{eq:box}
\ee
The box function in (\ref{eq:wppi}) takes into account that pions are 
produced at the beginning of the decay tunnel and that undecayed pions 
are absorbed by the wall at the end of the tunnel. 
The pion is assumed to have been produced at the time $t=0$, 
and the factor $e^{-\Gamma t/2}$ where $\Gamma$ is the pion decay rate 
in the laboratory frame takes into account the exponential decay of the 
pion's wave function. Obviously, the production at $t=x=0$ is an 
approximation, as the pion production process has finite space-time 
extension. Strictly speaking, eq.~(\ref{eq:wppi}) is only valid outside 
the pion production region, whose size we neglect here. 

In general, when the spatial width of the pion wave packet $\sigma_{x\pi}$ 
is considered to be finite, one can also employ the pion wave packet of the 
type (\ref{eq:wppi}) with $\delta(x-v_\pi t)$ replaced by the corresponding 
finite-width shape factor $g_\pi(x-v_\pi t)$, provided that the effect of 
the pion production process itself on neutrino oscillations can be 
neglected. This approximation is valid when the spatial width of the pion 
wave packet is negligibly small compared to the neutrino oscillation 
length $l_{\rm osc}$ and, in addition, the pion decay effects during 
its formation time are negligible: 
\be
\Gamma \sigma_{x\pi}/v_\pi \ll 1\,.    
\label{eq:cond}
\ee
We shall consider the case of finite $\sigma_{x\pi}$ in sections  
\ref{sec:finlength} and \ref{sec:altern}.

Throughout most of this paper, we shall be assuming that the muon 
produced alongside the neutrino in the pion decay is undetected and 
that its possible interaction with the environment can be ignored. 
In this case the muon is completely delocalized ($\sigma_{x\mu}\to \infty$) 
and therefore can be described by a plane wave \cite{AK}. 
The effects of possible muon detection or interaction with medium 
will be considered in section \ref{sec:mupi}.

\subsection{\label{sec:wp}Calculation of the neutrino wave packet
\label{sec:nuwp}}

To find the neutrino wave packet, we first calculate the amplitude 
${\cal A}(\pi\to \mu \nu_j)$ of the pion decay with the production of a
mass-eigenstate neutrino $\nu_j$ of mass $m_j$. We will be assuming that 
the pion is  described by the wave function (\ref{eq:wppi}), while 
the muon and the neutrino are described by plane waves of momenta $K$ 
and $p$, respectively. This will give us the probability amplitude that 
$\nu_j$ and the muon are produced with the momenta $p$ and $K$.  
The standard calculation gives 
\be
f_j^{S}(p,K)=
M_P\int_{-\infty}^\infty \!\!dt \!\int_{-\infty}^{\infty}\!dx \;
e^{i [E_j(p)+E_\mu(K)] t-i (p+K) x}\,
\psi_\pi(x,t)\,.
\label{eq:fj1}
\ee
Here $E_j(p)=(p^2+m_j^2)^{1/2}$, $E_\mu(K)=(K^2+m_\mu^2)^{1/2}$, and $M_P$ is 
the coordinate-independent part of the pion decay amplitude. 
In obtaining (\ref{eq:fj1}) it was taken into account that
the pion decay amplitude $M_P(q,k)$ is a smooth function of the pion and muon 
momenta $q$ and $k$ and therefore it can be replaced by its value $M_P\equiv 
M_P(Q,K)$ taken at the mean momenta (for the muon, $k=K$). The leptonic mixing 
parameter $U_{\mu j}$ is not included in the $\nu_j$ production amplitude 
(\ref{eq:fj1}) since it will be explicitly taken into account in the 
definition of the oscillation probability. Though the spatial integration in 
(\ref{eq:fj1}) is formally performed in the infinite limits, the box function 
in the expression (\ref{eq:wppi}) for $\psi_\pi(x, t)$ implies that in reality 
it only extends over the interval $[0, l_p]$. From (\ref{eq:wppi}) it also 
follows that the integral over time in (\ref{eq:fj1}) receives non-zero 
contribution only from the region $t\ge 0$. 

For a fixed $K$ the quantity $f_j^S(p,K)$ gives the amplitude of the 
neutrino momentum distribution, i.e.\ the momentum-space wave packet 
$f_j^S(p)$ of the produced $\nu_j$.
Substituting (\ref{eq:wppi}) into (\ref{eq:fj1}) and performing the 
integrations, we obtain  
\be
f_{j}^S(p) =
C_j\, \frac{1-e^{i [E_j(p)-E_P-v_\pi(p-P)+i\Gamma/2]\,l_p/v_\pi }}
{E_j(p)-E_P-v_\pi(p-P)+i\Gamma/2}\,.
\label{eq:fj2}
\ee
Here $C_j$ is a constant, and the quantities $P$ and $E_P$ are defined 
as 
\be 
P\equiv Q-K\,,\quad 
E_P\equiv E_{\pi}(Q)-E_{\mu}(K)\,, 
\label{eq:enmom} 
\ee 
i.e.\ in the plane-wave limit they would be the momentum and energy of 
the emitted neutrino. The momentum distribution amplitude $f_j^S(p)$ 
contains the usual Lorentzian energy distribution factor corresponding 
to the decay of an unstable parent state. The second term in 
the numerator of (\ref{eq:fj2}) (the exponential phase factor) reflects 
the fact that the parent pion is not completely free but is confined to 
a tunnel of length $l_p$; it would be absent in the case of decay of free 
pions, which can be considered as the limit $l_p \to \infty$.

The oscillation amplitude ${\cal A}_{\alpha\beta}(L,t)$ can now be obtained 
as a projection of the evolved neutrino state onto the detected one 
directly in the momentum space.  
However, the coordinate-space approach is more illuminating, and therefore 
we present it first. Momentum-space calculations will be employed in 
section \ref{sec:finlength} for the case $\sigma_{x\pi}\ne 0$.

\subsection{Neutrino wave packet in the coordinate space 
\label{sec:coord}}

The neutrino wave function in the coordinate space $\psi_j^S(x,t)$ can be 
obtained by Fourier-transforming the momentum-space neutrino wave packet 
(\ref{eq:fj2}):
\be
\psi_j^S(x,t)=\int\frac{dp}{2\pi}\, f_j^S(p) e^{-iE_j(p)t+i p x}\,.
\label{eq:psij1}
\ee
Substituting (\ref{eq:fj2}) into 
(\ref{eq:psij1}), expanding $E_j(p)$ near the point $p=P$ up to terms linear 
in $p-P$ and performing the integration, we find
\be
\psi_j^S(x,t)=const.\, e^{-i E_j(P_j)t+iP_j x}
\Big\{
e^{-\frac{\Gamma}{2(v_j-v_\pi)}(v_j t-x)}
\,{\rm box}\big(v_j t-x;\,\frac{v_j-v_\pi}{v_\pi}l_p,\,0 \big)\Big\}\,.
\label{eq:psij2}
\ee
Here the following notation has been used:
\be
P_j \equiv P+\frac{E_P-E_j(P)}{v_j-v_\pi}\,,
\qquad \mbox{where}\qquad
v_j\equiv \frac{\partial E_j(p)}{\partial 
p}\big|_{p=P}=\frac{P}{E_j(P)}\,,
\label{eq:Pj}
\ee
\be
E_j(P_j)\simeq E_j(P)+
v_j (P_j-P)
=E_j(P)+v_j\frac{E_P-E_j(P)}{v_j-v_\pi}\,.
\qquad\qquad~\,
\label{eq:Ej}
\ee
Note that for finite $l_p$ the momentum distribution $|f_j^S(p)|^2$ is 
an oscillating function of $p$, and $p=P_j$ corresponds to the peak 
of the envelope of this function. 

\begin{figure}
  \begin{center}
\includegraphics[width=6.0cm]{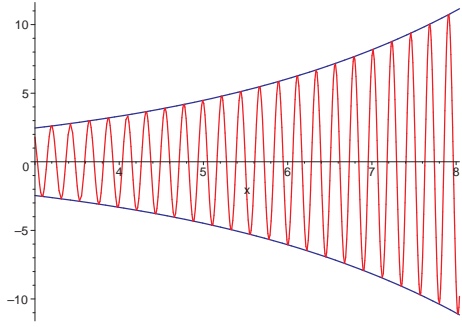}
  \end{center}
\caption{Schematic representation of the coordinate dependence 
of the neutrino wave packet (\ref{eq:psij2}) (arbitrary units).}
  \label{fig:wp}
\end{figure}

The wave packet in eq.~(\ref{eq:psij2}) has the general form 
(\ref{eq:wpgen}) with the shape-factor function 
\be
g(x-v_j t) = 
\Big\{
e^{-\frac{\Gamma}{2(v_j-v_\pi)}(v_j t-x)}
\,{\rm box}\big(v_j t-x;\,\frac{v_j-v_\pi}{v_\pi}l_p,\,0 \big)\Big\}\,.
\label{eq:gfact}
\ee
The presence of the box function here means that the neutrino wave 
packet has sharp front and rear edges %
\footnote{Note that this is related to our assumption of pointlike 
pions. For pion wave packets of finite spatial size (and e.g.\ Gaussian 
form) the neutrino wave packets would not have sharp borders.}. 
The box function enforces that the front of the neutrino wave packet arrives 
at the point 
$x$ at the time $t_1=x/v_j$
and leaves this point at $t_2=x/v_j+(1/v_\pi-1/v_j)l_p$.%
\footnote{ 
This has a simple interpretation: the emission of neutrino wave packet 
by the pion starts at $t=0$ and abruptly ends when the pion reaches the 
end of the decay tunnel, i.e.\ at $t=l_p/v_\pi$. Therefore $t_1=x/v_j$ 
and $t_2=l_p/v_\pi+(x-l_p)/v_j=x/v_j+(1/v_\pi-1/v_j)l_p$.} 
The distance between the edges of the wave packet is thus  
$v_j(1/v_\pi-1/v_j)l_p$. 
The shape factor reaches its maximum at the front of the wave 
packet and decreases exponentially towards its end. This gives an 
interesting example of an asymmetric coordinate-space wave packet
(see fig.~\ref{fig:wp}).

Having found the expression for the neutrino wave function, we can now 
calculate the expectation value $\bar{x}$ of the neutrino coordinate 
as well as the coordinate dispersion, which gives the spatial width of the 
neutrino wave packet $\sigma_{x\nu}$. In the limit $\Gamma l_p/v_\pi \gg 1$, 
when the length of the decay tunnel $l_p$ is large compared to the pion decay 
length $l_{\rm decay}=v_\pi/\Gamma$, we find 
\be
\bar{x}\approx v_jt-\frac{v_j-v_\pi}{\Gamma}\,, \qquad 
\sigma_{x\nu} \approx \frac{v_j-v_\pi}{\Gamma}
\,.
\label{eq:lim1}
\ee 
Note that in this limit the box function 
in eqs.~(\ref{eq:psij2}) and (\ref{eq:gfact}) has to be replaced by 
$\theta(v_j t-x)$, where $\theta(x)$ is the Heaviside step function. 

In the opposite limit, $\Gamma l_p/v_\pi \ll 1$, we find 
\be
\bar{x}\approx v_jt-\frac{v_j-v_\pi}{2 v_\pi}l_p\,, \qquad
\sigma_{x\nu} \approx \frac{1}{2\sqrt{3}}\frac{v_j-v_\pi}{v_\pi}l_p\,.
\label{eq:lim2}
\ee 
Notice that in this limit only a small fraction of pions decays before 
being absorbed by the wall at the end of the decay tunnel. 

Although, as was mentioned above, the distance between the sharp borders of 
the neutrino wave packet is $[(v_j-v_\pi)/v_\pi]l_p$, the effective spatial 
width $\sigma_{x\nu}$ of the wave packet is not necessarily determined by 
this 
quantity: this is only the case when the exponential decrease of the 
envelope function is relatively slow, $\Gamma l_p/v_\pi \ll 1$ (see 
(\ref{eq:lim2})). In the opposite case, $\Gamma l_p/v_\pi \gg 1$, which 
corresponds to the decay of unconfined free pions, it is this exponential 
decrease that dominates $\sigma_{x\nu}$, as described by eq.~(\ref{eq:lim1}).  
It is interesting that the neutrino wave packets are of macroscopic size 
in the case we consider.

Next, we have to find the amplitude ${\cal A}_j(L,t)$ which is the 
contribution of the $j$th neutrino mass eigenstate to the transition 
amplitude ${\cal A}_{\alpha\beta}(L,t)$.  
In the coordinate-space calculation it is simply given by the projection of 
$\psi_j^S(x,t)$ onto the detected neutrino state $\psi_j^D(x)$:
\be
{\cal A}_j(L,t)\equiv\int dx\, \psi_j^{D*}(x)\, \psi_j^S(x,t)\,.
\label{eq:A2}
\ee
The detected state is in general a wave packet centered on the point $x=L$. 
In what follows we will be assuming that the detection process is well 
localized both in space and time, so that the wave function of the detected 
neutrino state can be represented in the configuration space by $\delta(x-L)$. 
This requires some explanations. The momentum-space wave function of 
the detected neutrino state $f_j^D(p)$ is given by a formula similar to that
in eq.~(\ref{eq:fj1}), with the integration over the detection coordinate 
$(t',\,x')$. We assume that the detection process is perfectly 
localized in space and time, so that the integrand of this formula contains 
the factor $\delta(x'-x_D) \delta(t'-t_D)$. As a result, $f_j^D(p)$ is actually 
momentum independent, and for the coordinate-space wave function 
of the detected neutrino state centered on the point $x=L$, which is a 
time-independent Fourier transform of $f_j^D(p)$ \cite{AS1,AK}, we find
$\psi_j^D(x)=\delta(x-L)$. 
{}From~(\ref{eq:A2}) we then obtain 
\be
{\cal A}_j(L,t)=\psi_j^S(L,t)\,.
\label{eq:A2a}
\ee
The oscillation probability $P_{\alpha\beta}(L)$ is given, up to a 
normalization factor, by (see, e.g., \cite{AS1})
\be P_{\alpha\beta}(L)=\sum_{j,k}U^*_{\alpha j} 
U_{\beta j}^{} U_{\alpha k}^{} U_{\beta k}^*\,I_{jk}(L)\,, 
\label{eq:P1} 
\ee 
where 
\be I_{jk}(L)\equiv \int_{-\infty}^\infty dt\, {\cal A}_j(L,t) {\cal 
A}_k^*(L,t)\,. 
\label{eq:Ijk1} 
\ee
{}From eq.~(\ref{eq:A2a}) we then find 
\be
I_{jk}(L)=\int_{-\infty}^\infty\! dt\,\psi_j^S(L,t) \psi_k^{S*}(L,t)\,.
\label{eq:Ijk2}
\ee
Substituting here the expressions for $\psi_j^S(x)$ and $\psi_k^S(x)$ 
from (\ref{eq:psij2}) and performing the integration, we obtain 
\be
I_{jk}(L)=C_1
\cdot \frac{i\Gamma}
{v_\pi\frac{\Delta m_{jk}^2}{2P}+i\Gamma}
\left[
e^{-i\frac{\Delta m_{jk}^2}{2P}L}-e^{-\Gamma l_p/v_\pi}
e^{-i\frac{\Delta m_{jk}^2}{2P}(L-l_p)}
\right].
\label{eq:Ijk4}
\ee
Here we have discarded terms $\sim (\Delta m_{jk}^2)^2$ and 
$\sim \Delta m_{jk}^2 \Gamma/P$ and neglected the difference between $v_j$ 
and $v_k$ whenever they are multiplied by $\Gamma$ or by $\Delta m^2/2P$.%
\footnote{Note that the latter approximation implies that we neglect the 
effect of decoherence due to the wave packet separation (propagation 
decoherence). 
This is justified for $L\ll l_{\rm coh}$, where $l_{\rm coh}\simeq 
(v_g/|v_j-v_k|)\sigma_{x\nu}$, which is the case we are mainly interested 
in. The propagation decoherence effects can, however, be readily taken 
into account.}
The constant $C_1$ can be found by imposing the unitarity constraint 
$\sum_\beta P_{\alpha\beta}(L)=1$ on the oscillation 
probability (\ref{eq:P1}),%
\footnote{This normalization prescription looks rather {\em ad hoc} in the 
quantum-mechanical wave packet approach but can be rigorously 
justified in the quantum field theoretic framework \cite{AK}.} 
which leads to the normalization condition $I_{jj}(L)=1$ \cite{AS1}. 
This yields $C_1=(1-e^{-\Gamma l_p/v_\pi})^{-1}$, so that we finally 
obtain  
\be
I_{jk}(L)=\frac{1}{(1-e^{-\Gamma l_p/v_\pi})}
\cdot \frac{i\Gamma}
{v_\pi\frac{\Delta m_{jk}^2}{2P}+i\Gamma}
\left[
e^{-i\frac{\Delta m_{jk}^2}{2P}L}-e^{-\Gamma l_p/v_\pi}
e^{-i\frac{\Delta m_{jk}^2}{2P}(L-l_p)}
\right].
\label{eq:Ijk5}
\ee

By making use of the normalization condition $I_{jj}(L)=1$ one can 
demonstrate that the quantity $I_{jk}(L)$ 
(and therefore the 
oscillation probability (\ref{eq:P1})) is Lorentz invariant \cite{AS1}.

\subsection{\label{sec:prob}The oscillation probability}

With the expression for $I_{jk}(L)$ at hand, we can now calculate the 
oscillation probabilities from eq.~(\ref{eq:P1}). Consider a 2-flavour case 
when the oscillations are governed by just one mass squared difference and 
one mixing angle. As an example, short-baseline oscillations in the 3+1 
scheme with one sterile neutrino are essentially reduced to effective 
2-flavour ones with $\Delta m^2 \equiv \Delta m_{41}^2 \ne 0$, while  
the much smaller mass squared differences $\Delta m_{31}^2$ and 
$\Delta m_{21}^2$ can be neglected. In this case the survival probabilities 
$P_{\alpha\alpha}$ are described by the effective mixing parameters 
\be
\sin\theta\equiv s=|U_{\alpha 4}|\,,\qquad\quad \cos\theta\equiv c=
(1-|U_{\alpha 4}|^2)^{1/2}\,.
\label{eq:sc}
\ee
In the case of transition probabilities $P_{\alpha\beta}$ 
($\alpha \ne \beta$), the mixing parameter can be chosen as 
\be
\sin^2 2\theta \equiv 4|U_{\alpha 4}|^2 |U_{\beta 4}|^2\,.
\label{eq:s2}
\ee

Consider, for example, the survival probability of muon neutrinos in 
the 2-flavour scheme. From eqs.~(\ref{eq:P1}) and (\ref{eq:Ijk5}) we find  
\be
P_{\mu\mu}=c^4+s^4+\frac{2 c^2 s^2}{\xi^2+1}
\frac{1}{(1-e^{-\Gamma l_p/v_\pi})}
\left[
\cos \phi+\xi \sin\phi
-e^{-\Gamma l_p/v_\pi}[
\cos(\phi-\phi_p)+\xi
\sin(\phi-\phi_p)]
\right].
\label{eq:P2}
\ee
Here 
\be
\phi\equiv \frac{\Delta m^2}{2P}L\,,\qquad
\phi_p\equiv \frac{\Delta m^2}{2P}l_p \,,
\label{eq:phi}
\ee
and we have defined the parameter  
\be
\xi\equiv v_\pi \frac{\Delta m^2}{2P \Gamma}\,,
\label{eq:xi}
\ee
which, along with $\Gamma l_p/v_\pi$, characterizes possible decoherence 
effects at neutrino production  
(see the discussion in section \ref{sec:coh}). In the limit 
$v_\pi=1$ eq.~(\ref{eq:P2}) coincides with result found in ref.~\cite{HS}.

We shall now analyze the probability (\ref{eq:P2}) in the light of 
neutrino production coherence.

\subsection{\label{sec:coh}Coherence violation at neutrino production}  

Let us start with a qualitative analysis of neutrino production 
coherence. Recall that the production coherence condition requires 
that different neutrino mass eigenstates forming a flavor neutrino state be 
emitted coherently in the production process. This condition can be 
written as 
\be 
\Delta E \ll \sigma_E\,,
\label{eq:coher1}
\ee
where $\Delta E$ is the energy difference of 
different neutrino mass eigenstates and $\sigma_E$ is the quantum-mechanical 
energy uncertainty inherent to the production process. As was pointed 
out in \cite{AS1}, for decays of non-relativistic parent particles contained 
in a box of finite size, $\sigma_E$ is determined by the larger of the 
two quantities: the particle's decay width $\Gamma$ and the inverse time 
between two subsequent collisions of the particle with the walls of the box. 
In our case, the ``collision time'' is the time $l_p/v_\pi$ between the pion 
production and absorption of undecayed pions at the end of the decay tunnel. 
In general, for relativistic parent particles of velocity $v_P$ 
this non-relativistic energy uncertainty  
should be multiplied by the factor $v_g/(v_g-v_P)$. This takes into account 
that the neutrino is emitted by a moving particle in the forward direction, 
so that the parent particle ``chases'' the produced neutrino wave packet 
\cite{FS,AS1}. We thus find that in our case 
\be
\sigma_E\sim \max\{\Gamma, v_\pi/l_p\}
\frac{v_g}{v_g-v_\pi}\,.
\label{eq:sigmaE}
\ee

It is known that the energy uncertainty $\sigma_E$ also determines the spatial 
width of the neutrino wave packet \cite{AS1}: $\sigma_{x\nu}\simeq v_g/
\sigma_E$. {}From (\ref{eq:sigmaE}) we therefore find that in the limiting case 
$\Gamma l_p/v_\pi \gg 1$ the width of the wave packet $\sigma_{x\nu} \sim 
(v_g-v_\pi)/\Gamma$, whereas for $\Gamma l_p/v_\pi \ll 1$ one finds 
$\sigma_{x\nu} \sim [(v_g-v_\pi)/v_\pi]l_p$. These results are in full accord 
with our previously found values of $\sigma_{x\nu}$ (see eqs.~(\ref{eq:lim1}) 
and (\ref{eq:lim2})). 

As follows from 
(\ref{eq:Ej}), 
\be
\Delta E\equiv |E_j(P_j)-E_k(P_k)|
\simeq \frac{\Delta m^2}{2P}\frac{v_\pi v_g}{(v_g-v_\pi)}\,.
\label{eq:DeltaE}
\ee
Taking into account eq.~(\ref{eq:sigmaE}), for the (de)coherence parameter 
$\Delta E/\sigma_E$ we then find 
\be
\frac{\Delta E}{\sigma_E}\simeq \frac{\Delta m^2}{2P}l_p = \phi_p\,,
\quad(\Gamma l_p/v_\pi \ll 1)\,;\qquad\qquad
\frac{\Delta E}{\sigma_E}\simeq \xi 
\quad(\Gamma l_p/v_\pi \gg 1)\,.
\label{eq:Coher1}
\ee

Let us now consider the probability (\ref{eq:P2}) from the point of view 
of the production decoherence effects (see also \cite{HS}). As was mentioned 
above, these effects depend in general on two parameters, $\xi$ 
and $\Gamma l_p /v_\pi$. Note that these quantities have different dependence  
on the pion lifetime $\Gamma$, and that there is a relation between them  
and the phase $\phi_p$:
\be
\xi \cdot\frac{\Gamma l_p}{v_\pi}=\phi_p\,.
\label{eq:rel5}
\ee
The three quantities in this equation can also be expressed through  
the pion decay length $l_{\rm decay}=v_\pi/\Gamma$ and the neutrino 
oscillation length $l_{\rm osc}=4\pi P/\Delta m^2$:
\be
\xi = 2\pi\frac{l_{\rm decay}}{l_{\rm osc}}\,,\qquad\quad
\frac{\Gamma l_p}{v_\pi} = \frac{l_p}{l_{\rm decay}}\,,\qquad\quad
\phi_p = 2\pi\frac{l_p}{l_{\rm osc}}\,.
\label{eq:rel6}
\ee

Consider first the limit $\Gamma l_p/v_\pi \ll 1$. 
As follows from (\ref{eq:Coher1}), the decoherence parameter in this 
case is $\phi_p$. Thus, one can expect strong decoherence effects for 
$\phi_p \gg 1$ and no decoherence in the opposite limit.

The case $\Gamma l_p/v_\pi \ll 1$ is actually more easily studied by going 
to this limit in the expression for $I_{jk}(L)$ and then calculating the  
oscillation probabilities rather than by directly expanding the expressions 
for oscillation probabilities. Taking the limit $\Gamma l_p/v_\pi \ll 1$ in 
(\ref{eq:Ijk5}), we obtain
\be
I_{jk}(L)=\frac{i}{\frac{\Delta m_{jk}^2}{2P}l_p}\left[
e^{-i\frac{\Delta m_{jk}^2}{2P}L}-e^{-i\frac{\Delta m_{jk}^2}{2P}(L-l_p)}
\right].
\label{eq:Ijk6}
\ee
Note that this quantity satisfies the correct normalization condition  
$I_{jj}(L)=1$ provided that $\Delta m_{jj}^2$ is understood as the limit 
$\Delta m_{jk}^2 \to 0$. Substituting (\ref{eq:Ijk6}) into (\ref{eq:P1}), we 
find
\be
P_{\mu\mu}=c^4+s^4+\frac{2 s^2 c^2}{\phi_p}\big[\sin\phi 
-\sin(\phi-\phi_p)\big]\,. 
\label{eq:P3}
\ee
In the limit $\phi_p \gg 1$ this expression yields $P_{\mu\mu}\simeq 
c^4+s^4$, which corresponds to averaged neutrino oscillations, whereas 
for $\phi_p \ll 1$ 
it gives the standard oscillation probability 
\be
P^{\rm stand}_{\mu\mu}=c^4+s^4+2 s^2 c^2 \cos\phi\,,
\label{eq:Pstand}
\ee  
as expected. Let us stress that the latter result does not depend on whether 
$\xi$ is small or large, though $\xi$ must still satisfy the relation 
in eq.~(\ref{eq:rel5}). As follows from the same relation, in the case 
$\phi_p \gg 1$ (strong decoherence) $\xi$ is automatically very large in 
the $\Gamma l_p/v_\pi \ll 1$ regime.

Consider now  
the limit $\Gamma l_p/v_\pi \gg 1$, when most pions decay 
before reaching the end of the decay tunnel. In this case, 
according to (\ref{eq:Coher1}), decoherence effects should be governed by  
the parameter $\xi$. As expected, in the limit $\xi\to 0$ eq.~(\ref{eq:P2}) 
yields the standard oscillation probability (\ref{eq:Pstand}), whereas 
for $\xi \gg 1$ the oscillating terms in (\ref{eq:P2}) are suppressed 
and one obtains the averaged probability $P_{\mu\mu}\simeq c^4+s^4$.

Thus, we conclude that in the case of sufficiently small $l_p$, when both 
$\Gamma l_p/v_\pi\ll 1$ and $\phi_p \ll 1$, the production coherence 
condition is always satisfied (and the oscillation probability takes its 
standard form) irrespective of the value of $\xi$, which is irrelevant in 
this case. If $\Gamma l_p/v_\pi\ll 1$ but $\phi_p \gtrsim 1$ (which implies 
$l_{\rm osc}/2\pi \lesssim l_p \ll l_{\rm decay}$), the production coherence 
is moderately or strongly violated, depending on the value of $\phi_p$. 
In the case $\Gamma l_p/v_\pi\gg 1$ the production coherence depends on 
the parameter $\xi$: it is satisfied for $\xi \ll 1$ and strongly 
violated in the opposite case. 

Irrespectively of the value of $\Gamma l_p/v_\pi$, strong production 
coherence violation always implies $\xi \gg 1$, although large values of 
$\xi$ do not necessarily imply coherence violation. It is easy to see that 
the same applies to $\phi_p$. Therefore, in order to find out if the 
production coherence is violated one has to check the values of any two out 
of the three parameters $\Gamma l_p/v_\pi$, $\xi$ and $\phi_p$ (the third one 
will then be given by eq.~(\ref{eq:rel5})). It is convenient to choose 
these to be $\xi$ and $\phi_p$. 

Indeed, from the above considerations it is easy to see that if 
$\Gamma l_p/v_\pi$ is very small or very large, the production coherence is 
strongly violated in the case $\xi \gg 1$, $\phi_p \gg 1$ and is satisfied 
in all other limiting cases (large $\xi$ and small $\phi_p$, small $\xi$ 
and large $\phi_p$, and small $\xi$, small $\phi_p$). If 
$\Gamma l_p/v_\pi\sim 1$, the parameters $\xi$ and $\phi_p$ in the limiting 
cases are either both small or both large, and from eq.~(\ref{eq:P2}) 
it again follows that the production coherence is strongly violated if 
$\xi \gg 1$, $\phi_p \gg 1$. It is satisfied in the opposite case. 

In the intermediate cases one can expect moderate violation of the 
production coherence, which should lead to noticeable deviations of the 
oscillation probabilities from their standard (i.e.\ predicted under the 
coherence assumption) values.

\section{\label{sec:incoher}Incoherent probability summation approach}

Let us now consider oscillations of neutrinos produced in pion decays 
inside a decay tunnel in a entirely different approach \cite{HS}. We will 
be assuming that each individual neutrino production event is completely 
coherent, so that the oscillations of the produced neutrino are described 
by the standard probability (\ref{eq:Pstand}). At the same time, one has 
to take into account that the neutrino production can take place at 
any point along the decay tunnel, though the pion flux decreases 
exponentially with the distance from the pion production point. The 
oscillated neutrino flux at a detector at the distance $L$ from the 
beginning of the tunnel is then given by 
\be
F_{\mu}(E,L)\,=\, F_\pi(E,0)\Gamma \int_0^{l_p} e^{-\frac{\Gamma
x}{v_\pi}}\,P^{\rm stand}_{\mu\mu}(E, L-x) dx\,,
\label{eq:Fmu}
\ee
where $F_\pi(E,0)$ is the initially produced pion flux. The unoscillated 
neutrino flux $F_\mu^0$ (i.e.\ the $\nu_\mu$ flux in the absence of the 
oscillations) is given by eq.~(\ref{eq:Fmu}) with $P^{\rm stand}_{\mu\mu}
(E, L-x)$ in the integrand replaced by 1. One can now define the effective 
oscillation probability as the ratio of the oscillated and unoscillated fluxes: 
\be
P^{\rm eff}_{\mu\mu}(L,E) \;\equiv\;F_{\mu}(L,E)/
F_{\mu}^0(L,E)\,.
\label{eq:Peff}
\ee
The calculation is straightforward, and the result turns out to coincide 
exactly with the expression for $P_{\mu\mu}(E,L)$ given by eq.~(\ref{eq:P2}). 
Thus, the summation of the amplitudes of neutrino production at different 
points along the pion path performed in section \ref{sec:cohsum} and the  
summation of the corresponding probabilities carried out in this section 
lead to the same oscillation probability. We will discuss the reason for this 
intriguing coincidence and the conditions under which it is broken in section 
\ref{sec:altern} and in the Discussion section.

\section{\label{sec:finlength}Amplitude summation: The case of 
finite-width pion wave packets }
Let us now relax the assumption of pointlike pions 
that we employed in section~\ref{sec:cohsum}. For the pion wave 
function we will use the expression similar to (\ref{eq:wppi}), but 
with $\delta(x-v_\pi t)$ replaced by a general shape factor 
$g_\pi(x-v_\pi t)$: 
\be
\psi_\pi(x,t)=C_\pi\,e^{iQx-iE_\pi(Q)t-\Gamma
t/2}\,g_\pi(x-v_\pi t) \,{\rm box}(x; l_p, 0)\,.
\label{eq:wppi2}
\ee
We will be assuming that $g_\pi(x-v_\pi t)$ is peaked at or near the 
zero of its argument and rapidly decreases when $|x-v_\pi t|\gtrsim 
\sigma_{x\pi}$, but otherwise will not specify the exact form of this  
function. Substituting (\ref{eq:wppi2}) into (\ref{eq:fj1}), for 
the momentum-space wave packet of the $j$th mass eigenstate 
component of the produced neutrino we obtain 
\be
f_j^S(p)=M_P C_\pi \int_0^{l_p} dx\, 
e^{i[E_j(p)-E_P]\frac{x}{v_\pi}-i(p-P)x-\frac{\Gamma}{2 v_\pi}x}\; 
\bar{g}_\pi(E_j(p)-E_P)\,, 
\label{eq:fjk3}
\ee
where $\bar{g}_\pi(E_j(p)-E_P)$ is the Fourier transform of 
$g_\pi(x-v_\pi t)\,e^{-\frac{\Gamma}{2v_\pi}(v_\pi t-x)}$:
\be
\bar{g}_\pi(E_j(p)-E_P) = \int_{-\infty}^\infty dt'\,e^{i[E_j(p)-E_P] t'} \,
[g_\pi(-v_\pi t') e^{-\frac{\Gamma}{2}t'}]\,.
\label{eq:barg}
\ee
Here the integration variable is $t'=t-x/v_\pi$. Note that the integration 
has been extended to the negative semi-axis of $t'$  
here; this is justified only if for large negative $t'$ the function 
$g_\pi(-v_\pi t')$ decreases sufficiently rapidly, and in any case 
faster than $e^{-\Gamma|t'|/2}$. This condition 
is satisfied for both box-type and Gaussian pion wave packets which we 
will consider below.%
\footnote{There is an ambiguity in the way the decay exponential is 
introduced in the pion wave function, which is related  to the fact that 
for $\sigma_{x\pi}\ne 0$ one should take into account the finite duration 
$\sigma_{t\pi}$ of the pion production process. The decay exponential can 
be written as $e^{-\Gamma (t-t_0)/2}$, where $t_0$ is e.g.\ the time when 
the pion formation is completed, or the time when the pion production 
process starts. This ambiguity, however, plays no role if $\Gamma 
\sigma_{t\pi}\simeq \Gamma\sigma_{x\pi}/v_\pi \ll 1$, which we assume 
throughout this paper (see eq.~(\ref{eq:cond})). 
Note that the extra factor $e^{\Gamma t_0/2}$ can be absorbed into 
the normalization constant $C_\pi$ of the pion wave function and 
therefore does not affect the oscillation probabilities. }  

Performing the integration over $x$ in (\ref{eq:fjk3}), we obtain 
\be
f_j^S(p)=
C_j \frac{1-e^{i[E_j(p)-E_P-v_\pi(p-P)+
i\frac{\Gamma}{2}]\frac{l_p}{v_\pi}
}}
{E_j(p)-E_P-v_\pi(p-P)+i\frac{\Gamma}{2}}
\,\bar{g}_\pi(E_j(p)-E_P)\,,
\label{eq:fjk4}
\ee
where $C_j$ is a constant. This expression differs from the corresponding 
one in the case of pointlike pions (\ref{eq:fj2}) by the extra factor 
$\bar{g}_\pi(E_j(p)-E_P)$.

Let us now calculate the transition amplitude and the quantity $I_{jk}(L)$. We 
will be again assuming that the detection process is well localized in space 
and time. The contribution 
of the $j$th neutrino mass eigenstates to the oscillation amplitude is then 
given, according to eqs.~(\ref{eq:A2a}) and (\ref{eq:psij1}), by 
\be
{\cal A}_j(L,t)=\int \frac{dp}{2\pi}f_j^S(p) e^{ipL-iE_j(p)t}\,.
\label{eq:A4}
\ee 
Substituting this into eq.~(\ref{eq:Ijk1}) and performing the 
integration over $t$, we find 
\begin{align}
I_{jk}(L)=\int \frac{dp_1 dp_2}{(2\pi)^2}\,f_j^S(p_1)
f_k^{S*}(p_2)\,e^{i(p_1-p_2)L}\,2\pi\delta[E_j(p_1)-E_k(p_2)]
\nonumber \\
=e^{-i\frac{\Delta m_{jk}^2}{2P}L}\,\frac{1}{v_g}
\int \frac{dp_1}{2\pi}\,
f_j^S(p_1)f_k^{S*}(p_1+\Delta m_{jk}^2/2P)\,.
\label{eq:Ijk7}
\end{align}
Here we have used 
\be
\delta[E_j(p_1)-E_k(p_2)]\simeq \delta[v_g(p_1-p_2)+\Delta m_{jk}^2/2E]\,,
\label{eq:df}
\ee
which follows from $E_j(p_1)-E_k(p_2)\simeq 
(\partial E/\partial p)\Delta p +(\partial E/\partial m^2)\Delta m^2$. 
Substituting (\ref{eq:fjk4}) into (\ref{eq:Ijk7}), we obtain 
\begin{align}
I_{jk}(L)&=\frac{C_j C_k^*}{v_g^2}\, e^{-i\frac{\Delta 
m_{jk}^2}{2P}L}\,
\int \frac{dp}{2\pi}\,
\frac{[1-e^{i[E_j(p)-E_P-v_\pi(p-P)+
i\frac{\Gamma}{2}]\frac{l_p}{v_\pi}}]\bar{g}_\pi(E_j(p)-E_P)}
{E_j(p)-E_P-v_\pi(p-P)+i\frac{\Gamma}{2}}
\vspace*{2.5mm}
\qquad
\nonumber \\
&\times \frac{[1-e^{-i[E_k(p+\Delta m_{jk}^2/2P)-E_P-v_\pi(p+
\Delta m_{jk}^2/2P-P)-i\frac{\Gamma}{2}]\frac{l_p}{v_\pi}}]
\bar{g}_\pi^*\big(E_k\big(p+\frac{\Delta m_{jk}^2}{2P}\big)-E_P\big)}
{E_k\big(p+\frac{\Delta m_{jk}^2}{2P}\big)-E_P-v_\pi\big(p+\frac{\Delta 
m_{jk}^2}{2P}-P\big)-i\frac{\Gamma}{2}}\,.
\label{eq:Ijk8}
\end{align}
As usual, the constant $C_j$ should be found from the condition 
$I_{jj}(L)=1$. 

We shall now consider two special cases, Gaussian and box-type pion 
wave packets. 

\subsection{\label{sec:Gauss}Gaussian shape factor of the pion wave 
packet} 

In this case  
\be
g_\pi(x-v_\pi t)=e^{-\frac{(x-v_\pi t)^2}{4\sigma_{x\pi}^2}}\,,
\label{eq:Gauss}
\ee
which gives 
\be
\bar{g}(E_j(p)-E_P)=const.\, e^{-\frac{\sigma_{x\pi}^2}{v_\pi^2}[E_j(p)-E_P
+i\frac{\Gamma}{2}]^2}\,.
\label{eq:barg2}
\ee
Let us calculate $I_{jk}(L)$ in the limit $\Gamma l_p/v_\pi \gg 1$, which 
corresponds to the situation when most pions decay before they reach the 
end of the decay tunnel. Substituting (\ref{eq:barg2}) into (\ref{eq:Ijk8}) 
and expanding $E_j(p)$ near the point $p=P$ up to terms linear in $p-P$, 
we find 
\bea
I_{jk}(L)=C_{jk}\frac{e^{-i\frac{\Delta m_{jk}^2}{2P}L}}{1-i\xi}
\Big\{
e^{2\sigma_{x\pi}^2\Big(\frac{\Gamma}{2(v_g-v_\pi)}\frac{v_g}{v_\pi}+
i B_j\Big)^2}
{\rm erfc}
\Big[\sqrt{2}\sigma_{x\pi}\Big(\frac{\Gamma}{2(v_g-v_\pi)}\frac{v_g}{v_\pi}+
i B_j\Big) \Big]
\nonumber \\
+e^{2\sigma_{x\pi}^2\Big(\frac{v_g}{v_\pi}\frac{\Gamma}{2(v_g-v_\pi)}-
i B_k\Big)^2}
{\rm erfc}
\Big[\sqrt{2}\sigma_{x\pi}\Big(\frac{\Gamma}{2(v_g-v_\pi)}
\frac{v_g}{v_\pi}-i B_k\Big) 
\Big]\Big\}\,, 
\label{eq:Ijk7n}
\eea
where ${\rm erfc(z)}$ is the complementary error function, the constant 
$C_{jk}\propto C_j C_k^*$ is fixed by the condition $I_{jj}(L)=1$, and 
\be
B_j \equiv \frac{E_P-E_j(P)}{v_j-v_\pi}\,.
\label{eq:Bj}
\ee
The oscillation probabilities can now be obtained from eq.~(\ref{eq:P1}). 
We will consider the limit of small $\sigma_{x\pi}$, in which the result 
simplifies considerably. In the lowest non-trivial order in this 
parameter we obtain, after the proper normalization, 
\be
I_{jk}(L)=
\frac{e^{-i\frac{\Delta m_{jk}^2}{2P}L}}{1-i\xi}\Big[
1+i\frac{2}{\sqrt{2\pi}}\,\frac{v_g}{v_g-v_\pi}\,
\frac{\Delta m_{jk}^2}{2P}\,
\sigma_{x\pi}\Big]\,.
\label{eq:Ijk9}
\ee  
For the $\nu_\mu$ survival probability we then find 
\be
P_{\mu\mu}=c^4+s^4+\frac{2 c^2 s^2}{\xi^2+1}
\left[(\cos \phi+\xi \sin\phi)-A_\pi (\xi \cos\phi-\sin\phi)\right]\,,
\label{eq:Pmm1n}
\ee
where
\be
A_\pi = A_{\pi \rm Gauss}=\frac{2}{\sqrt{2\pi}}\,\frac{v_g}{v_g-v_\pi}\,
\frac{\Delta m_{jk}^2}{2P}\,\sigma_{x\pi}\,.
\label{eq:ApiGauss}
\ee
The term independent of $\sigma_{x\pi}$ here coincides with the 
$\Gamma l_p/v_\pi \gg 1$ limit of the $\nu_\mu$ survival probability 
(\ref{eq:P2}) obtained in the case of neutrinos produced in decays of 
pointlike pions, whereas the term proportional to $A_\pi$ gives the 
correction due to the finite size of the pion wave packets. Note that 
eq.~(\ref{eq:Pmm1n}) is actually the expansion of  
\be
P_{\mu\mu}=c^4+s^4+\frac{2 c^2 s^2}{\xi^2+1}
\left[\cos (\phi-A_\pi)+\xi \sin(\phi-A_\pi)\right]\,,
\label{eq:Pmm2n}
\ee
to the lowest non-trivial order in $A_\pi$, which shows that to  
this order the effect of the finite-size pion wave packets reduces to an 
additional oscillation phase proportional to $\sigma_{x\pi}$.

\subsection{\label{sec:box}Box-type shape factor of the pion wave packet} 

Consider now a box-type shape factor for the pion wave packet:
\be
g_\pi(x-v_\pi t)=box(v_\pi t-x, d, 0)\,.
\label{eq:gpi}
\ee
It corresponds to the pion wave packet of width $d$ and the initial 
condition that the pion starts arriving in the decay tunnel at $t=0$. From 
(\ref{eq:barg}) we have  
\be
\bar{g}_\pi(E_j(p)-E_P)=i\,\frac{1-e^{i[E_j(p)-E_P+i\frac{\Gamma}{2}]
\frac{d}{v_\pi}}}{E_j(p)-E_P+i\frac{\Gamma}{2}}\,.
\label{eq:barg3}
\ee
To facilitate comparison with the Gaussian pion wave packet case, it will 
be convenient for us to express the parameter $d$ through the effective  
spatial width of the pion wave packet $\sigma_{x\pi}$. The latter we always 
define as the coordinate dispersion in a given state, 
$[\langle x^2\rangle-\langle x \rangle^2]^{1/2}$.  For the pion state 
(\ref{eq:gpi}) we thus find 
\be
\sigma_{x\pi}=\frac{d}{2\sqrt{3}}\,. 
\label{eq:dsigma}
\ee

Consider once again the limit $\Gamma l_p/v_\pi \gg 1$. 
Substituting (\ref{eq:barg3}) into (\ref{eq:Ijk8}) gives, in the 
leading non-trivial order in $\sigma_{x\pi}$,  
\be
I_{jk}(L)=
\frac{e^{-i\frac{\Delta m_{jk}^2}{2P}L}}{1-i\xi}\Big[
1+i\frac{1}{\sqrt{3}}\,\frac{v_g}{v_g-v_\pi}\,
\frac{\Delta m_{jk}^2}{2P}\,
\sigma_{x\pi}\Big]\,.
\label{eq:Ijk10}
\ee  
The probability $P_{\mu\mu}$ is then again given by eq.~(\ref{eq:Pmm1n})
but now with 
\be
A_\pi=A_{\pi \rm box}=\frac{1}{\sqrt{3}}\,\frac{v_g}{v_g-v_\pi}\,
\frac{\Delta m_{jk}^2}{2P}\,\sigma_{x\pi}\,.
\label{eq:ApiBox}
\ee
Comparing eqs.~(\ref{eq:Ijk9}) and (\ref{eq:Ijk10}) we see that they have 
the same structure: there is a term of the zeroth order in $\sigma_{x\pi}$ 
which coincides with the corresponding expression in the 
$\Gamma l_p/v_\pi \gg 1$ limit of the case of pointlike pions (see 
eq.~(\ref{eq:Ijk5})), and the term 
which is linear in $\sigma_{x\pi}$. The coefficients in front of 
$\sigma_{x\pi}$ in the cases of box-type and Gaussian type pion wave 
packets differ only by a numerical factor. The same applies to the 
expression for $P_{\mu\mu}$ -- in both cases it is given by the same 
eq.~(\ref{eq:Pmm1n}) with the coefficients $A_\pi$ only differing by 
a numerical factor. 
Note that expressions (\ref{eq:Ijk9}), (\ref{eq:Ijk10}) and (\ref{eq:Pmm1n}) 
were obtained in the limit of small $A_\pi$ and therefore cannot describe 
neutrino production decoherence caused by the finite size of the pion wave 
packet (which would correspond to 
$A_\pi \gtrsim 1$); they just describe the appearance of extra oscillatory 
terms in the transition and survival probabilities rather than suppression of 
such terms. We will consider the decoherence effects due to $\sigma_{x\pi} 
\ne 0$ in section \ref{sec:largesigma}. For now, we will explain 
qualitatively the structure of the order $\sigma_{x\pi}$ correction term in 
the oscillation probability (\ref{eq:Pmm1n}) and in particular its 
proportionality to $v_g/(v_g-v_\pi)$. 

\subsection{\label{sec:qualit}Small corrections due to $\sigma_{x\pi}\ne 0$: 
a qualitative analysis}

If the pion wave packet has a finite spatial width, there is a an 
additional contribution $\Delta\phi$ to the oscillation phase stemming from 
the fact that the neutrino production region (and the space-time interval 
of integration over the neutrino production coordinate)  
is different in this case. 
One has therefore to add $\Delta \phi$ to the oscillation 
phase $\phi$ corresponding to the case of pointlike pions.

Neutrino emission in a decay of an extended pion is schematically 
illustrated in fig.~2. The wave packet of the pion is represented by a band 
in the $(t, x)$ plane, the muon wave packet is not shown. The slanted line 
represents the neutrino trajectory corresponding to the exact space and time 
localization of its detection. 
For neutrino emission 
from the space-time point $A$  the additional oscillation phase as 
compared to neutrino emitted from the point $B$ is given by the difference 
of the phases acquired by the mass eigenstates $\nu_j$ and $\nu_k$ along the 
segment $AB$ shown in the figure:
\be
\Delta \phi = -[E_j(P_j)-E_k(P_k)]\Delta t + (P_j-P_k)\Delta x\,.
\label{eq:phase1}
\ee
The quantities $\Delta t$ and $\Delta x$ are given by the projections 
of the segment $AB$ on the $t$ and $x$ axes. Simple geometric 
considerations yield  
\be
\Delta t =\frac{\sigma_{x\pi}}{v_g-v_\pi}\,, \qquad\quad 
\Delta x =\sigma_{x\pi}\frac{v_g}{v_g-v_\pi}\,.
\label{eq:deltas}
\ee
Substituting (\ref{eq:deltas}) into (\ref{eq:phase1}) and using 
eqs.~(\ref{eq:Pj}) and (\ref{eq:Ej}), we find 
\be
\Delta \phi \simeq -\frac{v_g}{v_g-v_\pi}\!\cdot\!\frac{\Delta m_{jk}^2}{2P} 
\sigma_{x\pi}\,.
\label{eq:phase2}
\ee
\begin{figure}
  \begin{center}
\includegraphics[width=8.0cm]{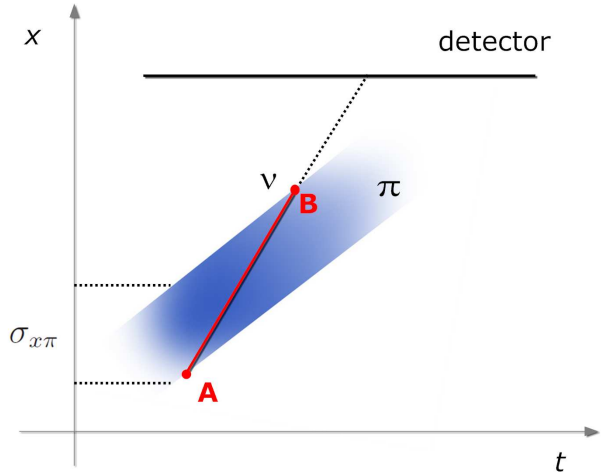}
  \end{center}
\caption{Schematical illustration of neutrino emission in pion 
decay in the case of finite-width pion wave packets.}
  \label{fig:scheme}
\end{figure}
This quantity does not depend on $t_A$ and therefore 
the same additional phase appears for neutrinos emitted at any time. 
The quantity $I_{jk}(L)$ can now be obtained from the one in the case 
of pointlike pions (eq.~(\ref{eq:Ijk5})) by substituting $\phi\to\phi+
\Delta \phi$. Taking the limit $\Gamma l_p/v_\pi \gg 1$ and keeping only 
the terms up to the first power in $\Delta \phi$, for the probability 
$P_{\mu\mu}$ we then obtain eq.~(\ref{eq:Pmm1n}) with $A_\pi=-\Delta
\phi$ (cf. eqs.~(\ref{eq:Pmm1n}) and (\ref{eq:Pmm2n})). The correction to 
the oscillation probability due to $\sigma_{x\pi} 
\ne 0$ obtained here from simple qualitative considerations coincides, up to 
numerical factors of order one, with those found in the previous 
subsections for the Gaussian and box-type pion wave packets. The exact 
value of the numerical coefficient depends on the shape of the pion wave 
packet. 

\section{\label{sec:thick}Probability summation and finite-thickness 
targets}

We shall now study the case when a proton beam is incident on a target 
of finite thickness and will treat this case in the framework of 
incoherent probability summation. 
This procedure is just the one that is usually employed in the analyses of 
neutrino oscillation experiments in order to take into account the extended 
sizes of neutrino sources (and detectors). The results of this section   
will also turn out to be useful for comparing the probability summation 
approach with the coherent amplitude summation in the case of finite widths 
of the pion wave packets. We therefore examine this case in detail. 

Consider a bunch of protons of 
duration $t_{bunch}$ with certain time distribution of protons within the 
bunch corresponding to a flux density $F_p (t)$. The protons are incident on 
a target of width $D$ and density distribution $n_{tar}(x)$ which in general 
depends on the coordinate $x$ along the direction of the beam. We assume that 
the protons produce pions on different scatterers incoherently.  
Consider the situation when the time of the pion production as well 
as the time of neutrino detection are not measured and the neutrino 
signal is summed up over the bunch. 
The total number of pions produced in the layer $(x,\,x + dx)$ of the 
target by the proton bunch is 
\be
d n_\pi (x)  =   k_\pi \int dt\, A(x) F_p(t)\, \sigma\, n_{tar}(x) dx 
= 
k_\pi \sigma F_p^{int} A(x) n_{tar}(x)  dx\,,  
\ee
where $\sigma$ is the $pN-$cross-section,  $A(x)$ is the factor describing 
the attenuation of the proton beam in the target by the time it reaches the 
point with the coordinate $x$, $k_\pi$ is the average number of pions 
produced in each interaction, and 
\be
F_p^{int} \equiv \int dt F_p(t) 
\ee
is the integral proton flux. Notice that at this point the time dependence 
factorizes out and disappears from the whole picture: obviously, the 
obtained result does not depend on a specific time dependence of the 
proton flux. 

The number of muon neutrinos at the detector 
which originate from the decays of pions born in the layer $(x, x+dx)$ of 
the target can be written as  
\be
dn_\pi (x) \int_{x}^{l_p} dx' e^{- \frac{\Gamma}{v_\pi} (x' - x)} 
P^{\rm 
stand}_{\mu\mu}(L - x')\,. 
\ee
Here $x'$ is the distance from the front end of the target (from the 
side 
of the incoming proton beam) to the point 
of the pion decay, $L$ is the distance from the target's front end to the 
detector, and $P^{\rm stand}_{\mu\mu}(L - x)$ is given by  
eqs.~(\ref{eq:Pstand}) and (\ref{eq:phi}) with $L$ replaced by $L-x$. 
The total number of muon neutrinos at the detector site, which 
originate from the decays of pions produced over the whole target, is  
\be
N_\nu = \sigma F_p^{int}  k_\pi  \int_0^{D} dz\,  n_{tar}(z) A(z) 
\int_{z}^{l_p} dx e^{-\frac{\Gamma}{v_\pi} (x - z)} 
P_{\mu\mu}^{\rm stand}(L - x)\,. 
\label{eq:nuflux}
\ee
Let us consider some special limits of this expression.\\ 

{1. Very thin target:} 
\be
n_{tar}(z) = n_0 \delta (z), ~~~ A(0) = 1. 
\ee
In this case 
\be
N_\nu = F_p^{int}  k_\pi  n_0 
\int_{0}^{l_p} dx e^{-\frac{\Gamma}{v_\pi} x} P_{\mu\mu}^{\rm stand}
(L - x)\,,
\ee
and the integration leads to the effective oscillation probability 
$P_{\mu\mu}$ that coincides with the one in eq.~(\ref{eq:P2}) (see 
(\ref{eq:Fmu})).

\bigskip
{2. Short decay tunnel:} $l_p \ll v_\pi/\Gamma$. 
In this case 
\be
N_\nu \approx F_p^{int}  k_\pi  \int_0^{D} dz ~ n_{tar}(z) A(z)
\int_{z}^{l_p} dx P_{\mu\mu}^{\rm stand}(L - x)\,.
\ee
Assuming for simplicity that $n_{tar}(z) = n_{tar} = const$ and   
$A(z) = A_0 = const$, we find 
\be
N_\nu \approx F_p^{int}  k_\pi n_{tar} A_0 
\int_0^{D} dz \int_{z}^{l_p} dx P_{\mu\mu}^{\rm stand}(L - x)\,.
\ee
Straightforward integration yields 
\be
P_{\rm eff}\equiv \frac{N_\nu}{N_\nu^0} \approx c^4 + s^4 + 2 c^2 s^2 
\frac{1}{\phi_p} \Big[\sin \phi  - \sin (\phi  - \phi_p) + 
\frac{D}{2 l_p} \left( \sin \phi - \sin (\phi - \phi_p) \right) - 
\frac{\pi D}{l_{\rm osc}} \cos \phi \Big],
\label{eq:probz}
\ee
where we have retained only terms up to the first order in $D$.  
In the limit $D \rightarrow 0$ this expression reduces to the one 
in eq.~(\ref{eq:P3}). 

{3. Extended target} and well localized protons. 
Assume that the decay tunnel is long, so that 
$l_{\rm decay}=v_\pi/\Gamma \ll l_p$. This is the same limit as the one used 
in the coherent amplitude summation calculations with finite-width pion wave 
packets carried out in sections \ref{sec:Gauss} and \ref{sec:box}.
We also assume that the density of the target is constant and that 
there is no absorption, i.e. $n_{tar} (z) A(z) = const$. Thus, we 
have the setup which corresponds to the one that in the amplitude summation 
approach led to the box-type pion wave packets. From eq.~(\ref{eq:nuflux}) 
we find 
\be
N_\nu \propto  \int_0^{D} dz 
\int_{z}^{L} dx e^{-\frac{\Gamma}{v_\pi}(x - z)} 
P_{\mu\mu}^{\rm 
stand}(L - x)\,.
\label{eq:nuflux1}
\ee
The integration yields  
\be
P_{\rm eff} = \frac{N_\nu}{N_\nu^0} 
\approx c^4 + s^4 + \frac{2 c^2 s^2}{1 + \xi^2}
\Big[\cos \phi + \xi \sin \phi 
- \frac{\Delta m^2}{4P}D  (\xi \cos \phi - \sin \phi) \Big], 
\label{eq:probz3}
\ee
where only corrections of the first order in $D$ are kept. 

Now let us compare the above results with those of the coherent amplitude 
summation approach with finite spatial width of the pion wave packets. 

What could be an analogue of the finite-width pion wave packet in the 
case when one performs the summation of contributions of neutrino production 
at different points at the level of probabilities? In that case the standard 
oscillation probability is used, and the notion of the pion wave packet 
does not apply. Nonetheless, one can introduce an analogy between the 
incoherent probability summation approach and the coherent amplitude 
summation with finite $\sigma_{x\pi}$ in the following way. 

In sec. \ref{sec:cohsum}, when 
considering the oscillations of neutrinos produced in pion decays in the 
amplitude summation approach, we estimated the width of the pion wave 
packets assuming that the pions are produced by proton collisions with 
nuclei of a target and that these nuclei are well localized and are 
essentially pointlike, whereas the protons are described by extended wave 
packets (see eqs.~(\ref{eq:epi}) and (\ref{eq:sigmaxpi})). However, one can 
instead imagine that the protons are pointlike, whereas the nuclei of the 
target are localized in relatively large spatial region of size $D$, 
so that their wave packets are characterized by the spatial width 
$\sigma_{xN}\simeq D$. In this case eqs.~(\ref{eq:rel01}), (\ref{eq:rel02}) 
and (\ref{eq:epi}) yield $\sigma_{xP}=\sigma_{xp}\approx 0$, $v_P\approx 
v_p$, $\sigma_{e\pi}\approx v_p/D$, and 
from eq.~(\ref{eq:rel0}) we find, instead of (\ref{eq:sigmaxpi}), 
\be
\sigma_{x\pi}\simeq \frac{v_p-v_\pi}{v_p}\,D\,.
\label{eq:sigmax}
\ee
On the other hand, in the probability summation approach one can identify the 
spatial size of the localization region of the target nuclei with the 
extension of the target in the direction of the proton beam. Therefore, 
when one employs the probability summation approach, the role of an analogue 
of the finite-size pion wave packets could be played by the 
finite-thickness target, with the correspondence between the thickness of 
the target $D$ in the probability summation approach and the width of 
the pion wave packet $\sigma_{x\pi}$ in the amplitude summation one 
established by eq.~(\ref{eq:sigmax}). 
Such a calculation was carried out in example 3 above, and the result 
is given in eq.~(\ref{eq:probz3}). 
Note that the obtained expression for the effective muon neutrino survival 
probability $P_{\rm eff}$ has the same structure as the probability 
$P_{\mu\mu}$ in eq.~(\ref{eq:Pmm1n}), which was derived in the coherent 
amplitude summation approach in the leading non-trivial order in 
$\sigma_{x\pi}$. The two expressions, however, differ by the coefficients in 
front of the second terms in the square brackets: while in the amplitude 
summation framework with box-type pion wave packets it was given by 
$A_{\pi{\rm box}}$ of eq.~(\ref{eq:ApiBox}), in the  probability summation 
approach of this subsection it is
\be
\frac{\Delta m^2}{4P}D = \frac{\Delta m^2}{4P}
\frac{v_p}{v_p-v_\pi}\,
\sigma_{x\pi}\,.
\label{eq:compare}
\ee
Here in the last equality we have used relation (\ref{eq:sigmax})
which establishes the correspondence between the proton target thickness 
$D$ of the probability summation approach and the width of the 
pion wave packet of the amplitude summation one. It is interesting to 
note that the expressions in eqs.~(\ref{eq:ApiBox}) and (\ref{eq:compare})
differ only by the replacement of the neutrino group velocity $v_g$ 
by the proton velocity $v_p$ and by the numerical factor $\sqrt{3}/2$. 

The fact that eqs.~(\ref{eq:Pmm1n}) and (\ref{eq:probz3}) have similar 
structure is actually not surprising: it can be shown under very general 
assumptions that the corrections due to the finite thickness of the 
proton target should have such a structure, irrespective of whether the 
coherent amplitude summation procedure or the incoherent probability 
summation one is employed. It should be stressed, however, that the 
coefficients of the correction terms are different. As will be shown in 
section \ref{sec:genform}, this is a reflection of the fact that the 
equivalence between the coherent and incoherent summation approaches in 
general only holds in the case of pointlike pions and localized in space and 
time neutrino detection. It can also be shown that the equivalence holds even 
in the case of extended pion wave packets provided that both the neutrino and 
muon detection processes are perfectly localized (see section 
\ref{sec:disc}).

\section{\label{sec:altern}An alternative approach to calculation of 
oscillation probabilities. Finite-width pion wave packets}

We shall now develop a different calculational approach to the 
coherent amplitude summation procedure, which is based on 
a different order of integrations involved in the calculation of 
$I_{jk}(L)$ and is valid for arbitrary shapes of the pion wave packets. 
This approach will be useful for explaining the equivalence of the results 
of the amplitude and probability summations of the contributions of neutrino 
production at different points, which was established above for the case of 
pointlike pions and perfectly localized neutrino detection. It will also be 
helpful for studying the cases when the muon emitted alongside the neutrino 
in pion decay is detected and when the interactions of pions in the bunch 
between themselves or with other particles inside the neutrino source are 
taken into account.  

\subsection{\label{sec:genform}General formalism}

We start with eq.~(\ref{eq:Ijk7}) for $I_{jk}(L)$ and substitute 
into it expression (\ref{eq:fj1}) for the momentum-space neutrino wave 
packet. The pion wave function is taken in the form~(\ref{eq:wppi2}) with 
an arbitrary shape factor $g_\pi(x-v_\pi t)$. The resulting expression 
contains integrations over two coordinates, two times and a momentum. 
Integrating first over the momentum, we obtain
\begin{align}
I_{jk}(L)=C_0 e^{-i\frac{\Delta m_{jk}^2}{2P}L}
&\!\int dx_1 dt_1\! \int dx_2 \,dt_2 
\;g_\pi(x_1-v_\pi t_1)\,g^*_\pi(x_2-v_\pi t_2)
e^{i[E_j(P)-E_P]t_1-i[E_k(P)-E_P]t_2}
\nonumber \\
&\times
e^{i\frac{\Delta m_{jk}^2}{2P}
[(x_1+x_2)-v_g(t_1+t_2)]-\frac{\Gamma}{2}(t_1+t_2)}
\,\delta[(x_1-x_2)-v_g(t_1-t_2)].
\label{eq:Ijk1n}
\end{align}
Here $C_0$ is a constant, and 
the integrations over $x_1$ and $x_2$ are performed in the interval 
$[0,\,l_p]$. The integrations over time are carried out over the whole 
interval $(-\infty, \infty)$, which is justified if $g_\pi(z)$ 
decreases sufficiently rapidly for large $|z|$ (see the discussion 
after eq.~(\ref{eq:barg})).  After the integration over 
the time variables, eq.~(\ref{eq:Ijk1n}) becomes 
\begin{eqnarray}
I_{jk}(L)=C e^{-i\frac{\Delta m_{jk}^2}{2P}L}\!\int dx_1\! \int dx_2 
\;e^{i\big[\frac{E_j(P)+E_k(P)}{2}-E_P\big]
\frac{(x_1-x_2)}{v_g}}\qquad
\nonumber \\
\times
e^{i\frac{\Delta m_{jk}^2}{2P}(x_1+x_2)-\frac{\Gamma}{2v_\pi}(x_1+x_2)}
G_\pi(x_1-x_2)\,,
\label{eq:Ijk2n}
\end{eqnarray}
where $C$ is a constant 
and the effective shape factor $G_\pi(x_1-x_2)$ is a convolution of two 
$g_\pi$ functions (taken at shifted arguments) with an exponential factor:  
\be
G_\pi(x_1-x_2) \equiv \!\int 
dy\,g_\pi\big(y+\frac{v_g-v_\pi}{2v_g}(x_1-x_2)\big)
\,g_\pi^*\big(y-\frac{v_g-v_\pi}{2v_g}(x_1-x_2)\big)\,
e^{\frac{\Gamma y}{v_\pi}}\,.
\label{eq:G}
\ee
The function $G_\pi(x_1-x_2)$ satisfies 
\be
G_\pi^*(z)=G_\pi(-z)\,.
\label{eq:G2}
\ee
If $g_\pi$ is real, then $G_\pi$ is also real and is an even 
function of its argument. 

Note that the shift of the arguments of the two $g_\pi$ functions in 
(\ref{eq:G}) is proportional to $(v_g-v_\pi)/v_g$. The presence of the two 
$g_\pi$ functions with shifted arguments in (\ref{eq:G}) is a reflection of 
the fact that we have a coherent summation of the amplitudes of 
neutrino production at different points here 
(there are interference terms between $g_\pi$ and $g^*_\pi$ which 
for the same value of $y$ have different arguments). For pointlike pions 
$g_\pi(x-v_\pi t)\propto 
\delta(x-v_\pi t)$, and instead of the product $g_\pi(z_1) g_\pi^*(z_2)$ we 
get $|g_\pi(z_1)|^2$, meaning that there is no interference terms and 
therefore the amplitude summation reduces to the probability summation. 
This gives the explanation of the equivalence of the two summation approaches, 
established in sections \ref{sec:cohsum} and \ref{sec:incoher}. We will 
discuss this point in more detail in the Discussion section.

If the pion shape factor $g_\pi(x-v_\pi t)$ is characterized by a width 
$\sigma_{x\pi}$ (i.e. $g_\pi(x-v_\pi t)$ quickly decreases when 
$|x-v_\pi t|$ becomes large compared to $\sigma_{x\pi}$), then, as follows 
from its definition, the 
function $G_\pi(x_1-x_2)$ is typically centered around the zero of its 
argument and is characterized by the width $[v_g/(v_g-v_\pi)]\sigma_{x\pi}$.
  
Equation~(\ref{eq:Ijk2n}) can be transformed into the form %
\footnote{
For this, one has to go to the integration variables $z\equiv x_1-x_2$ 
and $x\equiv \frac{x_1+x_2}{2}$, write the outer integral over $z$ as  
$\int_{-l_p}^{l_p}dz = \int_{-l_p}^0dz+\int_0^{l_p}dz$, change the 
integration variable in the first of these integrals according to $z \to 
-z$, and shift the integration variable in the inner integral over $x$  
according to $x \to x+z/2$.}
\begin{eqnarray}
I_{jk}(L)=C \int_0^{l_p} dz\, K_{jk}(z)\int_z^{l_p} dx\,
e^{-i\frac{\Delta m_{jk}^2}{2P}(L-x)-\frac{\Gamma}{v_\pi}(x-z)}\,,
\label{eq:Ijk3n}
\end{eqnarray}
where
\be
K_{jk}(z)\equiv 
\big[G_\pi(z) e^{i[E_k(P)-E_P]\frac{z}{v_g}}+
G_\pi^*(z) e^{-i[E_j(P)-E_P]\frac{z}{v_g}}\big] 
e^{-\frac{\Gamma}{2v_\pi}z}\,.
\label{eq:K1}
\ee
Eqs.~(\ref{eq:Ijk3n}) and (\ref{eq:K1}) are the final result of the 
calculational approach presented in this section. As usual, the constant 
$C$ in (\ref{eq:Ijk3n}) should be determined from the condition 
$I_{jj}(L)=1$. Note that, even though formally the integration over 
$z$ in (\ref{eq:Ijk3n}) extends from $0$ to $l_p$, in reality it is 
effectively limited from above by $z_{max}\sim [v_g/(v_g-v_\pi)]
\sigma_{x\pi}$ due to the properties of $G_\pi(z)$ (provided, of course, 
that $z_{max}<l_p$). For the special cases of Gaussian and box-type pion wave 
packets and $\Gamma l_p/v_\pi \gg 1$ eqs.~(\ref{eq:G}) - (\ref{eq:K1}) 
reproduce the results obtained 
in sections \ref{sec:Gauss} and \ref{sec:box}.   

It is interesting to note that there is a striking similarity between the 
general result (\ref{eq:Ijk3n}) for $I_{jk}(L)$ in the coherent amplitude 
summation approach and eq.~(\ref{eq:nuflux}) obtained for 
finite-thickness targets in the incoherent probability summation one. 
There would be a complete 
correspondence between the two results if the quantity $K_{jk}(z)$ in 
(\ref{eq:Ijk3n}) did not depend on the indices $j$ and $k$: $K_{jk}(z) \to 
K(z)$. In that case it would be possible to multiply 
the inner integral in~(\ref{eq:Ijk3n}) by $U_{\alpha j}^* U_{\alpha k} 
U_{\beta j} U_{\alpha k}^*$ and perform the summation over $j$ and $k$, 
which would produce $P_{\alpha\beta}^{\rm stand}(L-x)$ in the 
integrand. The same operation applied to the left-hand side 
of~(\ref{eq:Ijk3n}) yields, according to (\ref{eq:P1}), the oscillation 
probability $P_{\alpha\beta}(L)$. The correspondence between the two 
results would then be exact provided that one identifies $n_{tar}(z)
A(z)$ of eq.~(\ref{eq:nuflux}) with $K(z)$ of eq.~(\ref{eq:Ijk3n}). 
The quantity $K_{jk}(z)$ is independent of the indices $j$ and $k$, e.g., in 
the case of pointlike pions. Indeed, from (\ref{eq:G}) 
and (\ref{eq:K1}) we find that in this case $G_\pi(z)\propto \delta(z)$ and 
$K_{jk}(z)=K(z)=2G_\pi(z)$. As we already know, for pointlike pions the 
results of the amplitude and probability summation approaches indeed coincide.

\subsection{\label{sec:largesigma}Large $\sigma_{x\pi}$ and production 
decoherence}

If the spatial width of the pion wave packets $\sigma_{x\pi}$ is 
sufficiently large, one can expect production decoherence due to the 
averaging of the oscillation phase along the pion wave packet. In 
analogy with the parameters $\xi$ and $\phi_p$ in eq.~(\ref{eq:rel6}), the 
decoherence parameter in this case is expected to be  
\be
\phi_\pi \equiv \frac{\Delta m^2}{2P}\, 
\Delta x =2\pi 
\frac{
\Delta x}{l_{\rm osc}}\,,
\label{eq:phipi}
\ee
where the quantity $\Delta x$ defined in eq.~(\ref{eq:deltas}) is the 
spatial size of the region over which the additional averaging of the 
oscillation phase occurs. 
Note that the parameter $\phi_p$ coincides with the introduced earlier 
quantity $A_\pi$ up to a numerical factor of order one. 
Neutrino production decoherence due to large $\sigma_{x\pi}$ might be 
expected to manifest itself through the suppression of the oscillatory terms 
of the flavour transition probabilities by \mbox{$\phi_\pi$-dependent} 
factors that vanish in the limit $\phi_\pi\to \infty$.  
However, large-$\sigma_{x\pi}$ expansions of the oscillation probabilities 
in the cases of Gaussian and box-type pion wave packets, for which we 
have closed-form expressions, do not reveal any suppression of the 
oscillations due to \mbox{$\phi_\pi$-dependent} factors. How can this be 
understood? 

Let us first consider the limit $\Gamma l_p/v_\pi \gg 1$. 
As follows from eq.~(\ref{eq:Ijk3n}), $I_{jk}(L)\propto 1/(1-i\xi)$ in this 
case. The oscillatory terms of the probabilities $P_{\alpha\beta}(L)$ will 
then be in any case suppressed if \mbox{$\xi \gg 1$}. Let us now notice that 
$\xi \gg \phi_\pi$, which follows from (\ref{eq:cond}).\,%
\footnote{Except for pions of extremely high energies, which are currently 
inaccessible. Indeed, $\xi/\phi_\pi = (l_{\rm decay}/\sigma_{x\pi})
[(v_g-v_\pi)/v_g]$. The first factor on the right hand side of this 
equality is extremely large because the pion decay length 
$l_{\rm decay}=v_\pi/\Gamma$ is a macroscopic quantity, whereas the width of 
the pion wave packet $\sigma_{x\pi}$ is microscopic. 
The condition $\xi/\phi_\pi\lesssim 1$ therefore can only be satisfied 
for extremely small values of   
$v_g-v_\pi\approx 1-v_\pi$. From a similar argument it also follows that
 $\phi_p\gg \phi_\pi$. }
It should be stressed that 
(\ref{eq:cond}) is a necessary condition for not including the pion production 
process in the description of neutrino oscillations, which is assumed to be 
satisfied throughout this paper. Thus, in the limit of large $\phi_p$ we have 
$\xi \gg \phi_p \gg 1$, and the oscillations are suppressed. 

Similarly, in the opposite limit $\Gamma l_p/v_\pi \ll 1$ the oscillations 
are suppressed provided that $\phi_p \gg 1$ (see the discussion in section 
\ref{sec:coh}). Note that $\phi_p \gg \phi_\pi$ because $l_p \gg  
\sigma_{x\pi}$. 
Therefore, in the limit $\Gamma l_p/v_\pi \ll 1$ 
the oscillations will be suppressed for large $\phi_\pi$ because of 
$\phi_p \gg \phi_\pi \gg 1$.
It is easy to see that for large $\phi_\pi$ the oscillations are also 
suppressed in the case $\Gamma l_p/v_\pi \sim 1$. We conclude that in the 
case $\phi_\pi \gg 1$ the oscillations are quenched as expected, even  
though not directly by $\phi_\pi$-dependent factors.

\section{\label{sec:mupi}Effects of muon detection and pion collisions}

Up to now we have been considering only the situations when the muon 
produced together with the neutrino in the pion decay is neither 
directly detected nor interacts with the medium, and the pions in the 
bunch do not interact with each other or with other particles which 
may be present in the neutrino source. We shall now lift these 
restrictions. 

As in section~\ref{sec:cohsum}, 
we shall be assuming here that the parent pions can be considered as 
pointlike and are described by the wave function (\ref{eq:wppi}). 
Consider the case when the muon produced alongside the neutrino in the 
pion decay is ``measured'', either directly by a dedicated detector or 
through its interaction with the particles of the medium. In this case the 
muon is localized by the interaction and therefore it 
should be described by a wave packet rather than by a plane wave. 
We adopt the wave function of muons of the form (\ref{eq:wpgen}): 
\be
\psi_\mu(x,t)=e^{iKx-iE_\mu(K)t}\,g_\mu[(x-x_S)-v_\mu(t-t_S)]\,.
\label{eq:wpmu}
\ee
Here $g_\mu[(x-x_S)-v_\mu(t-t_S)]$ is the shape factor of the muon wave 
packet, which is determined by the muon detection process.  
 It is assumed to have a peak at the zero of its argument and 
decrease rapidly when the modulus of the argument becomes large compared 
to $\sigma_{x\mu}$, where $\sigma_{x\mu}$ is the spatial width of the muon 
wave packet. Our choice of the argument of $g_\mu$ corresponds to the 
initial condition that at the time $t=t_S$ the peak of the muon wave packet 
is at $x=x_S$. We shall choose $x_S$ to be the coordinate of the 
center of the muon wave packet at the neutrino production time. Since we 
assume the pions to be pointlike and the point $x_S$ should lie on the pion 
trajectory, $x_S$ and $t_S$ must be related through $x_S=v_\pi t_S$.   
Obviously, $x_S \le l_p$. 

Repeating essentially the same calculations as in section~\ref{sec:altern}, 
we arrive at a very simple and compact expression for $I_{jk}(L)$: 
\be
I_{jk}(L)=C_0 \int_0^{l_p}dx\,
\big|g_\mu\big((x-x_S)\frac{v_\pi-v_\mu}{v_\pi}\big)\big|^2 \,
e^{-i\frac{\Delta m_{jk}^2}{2P}(L-x)-\Gamma\frac{x}{v_\pi}}\,.
\label{eq:Ijk11}
\ee
The argument of the shape factor function $g_\mu$ 
here implies that the effective width of the muon wave packet 
that enters into eq.~(\ref{eq:Ijk11}) is actually 
\be
\tilde{\sigma}_{x\mu}\equiv\sigma_{x\mu}\frac{v_\pi}{v_\pi-v_\mu}\,.
\label{eq:eff}
\vspace*{-2.0mm}
\ee
Note that this expression has a structure analogous to that of $\Delta x$ 
in eq.~(\ref{eq:deltas}) and allows a simple geometric interpretation in 
terms of the plots similar to the one in fig.~\ref{fig:scheme}.

The calculation leading to (\ref{eq:Ijk11}) was carried out in the 
quantum-mechanical approach with coherent amplitude summation of 
the contributions of neutrino production at different points along the pion 
path. Nevertheless, in the course of the calculation one had to integrate 
an expression containing $|g_\pi(z)|^2$ rather than $g_\pi(z_1) g_\pi^*(z_2)$. 
As was stressed in section~\ref{sec:altern}, this reflects the fact that for 
pointlike pions and localized neutrino detection the results of the 
amplitude and probability summation approaches coincide. Thus, the 
equivalence of these two approaches in this case holds 
true irrespective of whether the charged lepton accompanying the neutrino 
production is detected or not. This is also confirmed by the presence 
of the squared modulus of $g_\mu$ in the integrand of eq.~(\ref{eq:Ijk11}).  
We will discuss how the charged lepton detection can affect the 
equivalence of the two approaches in the cases when either the pion 
is not pointlike or the 
neutrino detection process is delocalized in section~\ref{sec:disc}. 

Let us now study the result in eq.~(\ref{eq:Ijk11}) in several limiting 
cases. First, we note that in the limit 
$\tilde{\sigma}_{x\mu}\to \infty$, which 
corresponds to the plane-wave approximation for the muon employed 
in sections \ref{sec:cohsum} - \ref{sec:altern}, the shape factor of the 
muon wave packet $g_\mu \to const$, and we recover the previously found 
expression (\ref{eq:Ijk5}) for $I_{jk}(L)$ and therefore the previous results 
for the oscillation probabilities. 

Consider now the case 
$\tilde{\sigma}_{x\mu}\to \!\!\!\!\!\!\!/ ~\;\infty$. Eq.~(\ref{eq:Ijk11}) 
then actually describes oscillations of a ``tagged'' neutrino, i.e.\ of 
a neutrino produced together with the muon which was detected and whose 
production coordinate was found to be $x_S$ with the accuracy given by 
$\tilde{\sigma}_{x\mu}$. Let us first consider the limit 
$\tilde{\sigma}_{x\mu}\to 0$. In this case in the integrand of 
eq.~(\ref{eq:Ijk11}) the function $g_\mu \propto \delta(x-x_S)$, i.e.\ the 
muon detection exactly localizes the neutrino production point.
Eq.~(\ref{eq:Ijk11}) then yields 
\be
I_{jk}(L)=const.\, 
e^{-\Gamma\frac{x_S}{v_\pi}}\,
e^{-i\frac{\Delta m_{jk}^2}{2P}(L-x_S)}\,.
\label{eq:Ijk12}
\ee
The real exponential factor here just describes the depletion of the pion 
flux by the time they reach the point $x=x_S$; this factor can be absorbed 
in the overall normalization of $I_{jk}(L)$ which does not affect the 
transition probability. The usual normalization procedure yields  
$I_{jk}(L)=\exp[-i\frac{\Delta m_{jk}^2}{2P}(L-x_S)]$, which leads to the 
standard probability $P_{\alpha\beta}^{\rm stand}(L-x_S)$ that describes  
$\nu_\alpha \to \nu_\beta$ oscillations between the neutrino production 
point $x_S$ and the detection point $L$. The production decoherence effects 
are absent in this case.

Consider now the case when the effective spatial width of the muon wave 
packets $\tilde{\sigma}_{x\mu}$ is neither infinite nor vanishingly small. 
For definiteness, we shall consider Gaussian muon wave packets with 
the shape factor $g_\mu(x-v_\mu t)=\exp[-\frac{(x-v_\mu t)^2}
{4\sigma_{x\mu}^2}]$, which gives
\be
g_\mu\big((x-x_S)\frac{v_\pi-v_\mu}{v_\pi}\big)
=e^{-\frac{(x-x_S)^2}{4\tilde{\sigma}_{x\mu}^2}}\,. 
\label{eq:Gauss2}
\ee

Let us now consider $I_{jk}(L)$, as given by eq.~(\ref{eq:Ijk11}) with 
$g_\mu$ from (\ref{eq:Gauss2}), in some limiting cases of interest. 
First, for very large $\tilde{\sigma}_{x\mu}$ one expects to recover the 
results of section~\ref{sec:cohsum}, which were obtained for the plane-wave 
description of the muon, i.e.\ in the limit $\tilde{\sigma}_{x\mu}\to \infty$. 
Indeed, taking into account that $0\le x_S \le l_p$, its is easy to see that for $\tilde{\sigma}_{x\mu}\gg l_p$ the 
function $g_\mu$ in the integrand of (\ref{eq:Ijk11}) can to a very good 
accuracy be replaced by unity, so that we get
\be
I_{jk}(L)=C_0\int_0^{l_p}dx\,
e^{-i\frac{\Delta m_{jk}^2}{2P}(L-x)-\Gamma\frac{x}{v_\pi}}\,,
\label{eq:Ijk11a}
\ee
which is just the result found in section~\ref{sec:cohsum} where the muon 
was assumed to go ``unmeasured'' and described by a plane wave. Indeed, 
direct integration shows that (\ref{eq:Ijk11a}) leads to (\ref{eq:Ijk5}). 

Eq.~(\ref{eq:Ijk11a}) is also obtained in the cases when 
$\tilde{\sigma}_{x\mu}\lesssim l_p$, provided that the main contribution to 
the integral in (\ref{eq:Ijk11}) comes from a relatively small region $x 
\lesssim x_{c}$ with $x_c \ll l_p$, and in addition $\tilde{\sigma}_{x\mu}\gg 
x_c$. For instance, if $\Gamma l_p/v_\pi \gg 1$, the integrand of 
(\ref{eq:Ijk11}) is strongly suppressed for $x\gg v_\pi/\Gamma$ due to the 
exponential decay factor. Thus, for $\Gamma/v_\pi \gtrsim \Delta m_{jk}^2/2P$ 
we have $x_c\sim v_\pi/\Gamma$. In this case the factor $g_\mu$ in the 
integrand can be replaced by unity and the results of section~\ref{sec:cohsum} 
are recovered provided that 
\be
\tilde{\sigma}_{x\mu}\gg \max\Big\{\frac{v_\pi}{\Gamma},\, 
\Big(\frac{v_\pi}{\Gamma}x_S\Big)^{1/2}\Big\}.
\label{eq:cond2}
\ee

Analogously, if $(\Delta m_{jk}^2/2P)l_p \gg 1$ and $\Delta m_{jk}^2/2P
\gtrsim \Gamma/v_\pi$ we have $x_c\sim 2P/\Delta m_{jk}^2$ (for larger values 
of $x$ the integrand of (\ref{eq:Ijk11}) is fast oscillating and the 
corresponding contributions to the integral are strongly suppressed). In 
this case the results of section~\ref{sec:cohsum} are recovered provided that 
\be
\tilde{\sigma}_{x\mu}\gg \max\Big\{\frac{2P}{\Delta m_{jk}^2},\, 
\Big(\frac{2P}{\Delta m_{jk}^2} x_S\Big)^{1/2}\Big\}.
\label{eq:cond3}
\ee

Let us now consider the case of small $\tilde{\sigma}_{x\mu}$.
In the limit 
\be
\tilde{\sigma}_{x\mu}\ll \min\{l_p,\,x_S\}\,,
\label{eq:cond4}
\ee
one can extend the integration in (\ref{eq:Ijk11}) over the whole infinite 
axis of $x$. The integration with $g_\mu$ from~(\ref{eq:eff}) then gives, 
after the usual normalization procedure,  
\be
I_{jk}(L)\simeq e^{-i\frac{\Delta 
m_{jk}^2}{2P}(L-x_S+\frac{\Gamma}{v_\pi}\tilde{\sigma}_{x\mu}^2)}
\,e^{-\frac{1}{2} \big(\frac{\Delta m_{jk}^2}{2P}\big)^2
\tilde{\sigma}_{x\mu}^2}\,.
\label{eq:Ijk13}
\ee
Let us first look at the exponential phase factor in this formula.
In the absence of the second (real) exponential factor, it would just describe 
the standard $\nu_\alpha \to \nu_\beta$ oscillations between the points with 
the coordinates $x_S-(\Gamma/v_\pi)\tilde{\sigma}_{x\mu}^2$ and $L$. The term 
$(\Gamma/v_\pi)\tilde{\sigma}_{x\mu}^2$ in the coordinate of the initial point 
describes the shift of the position of the center of the neutrino production 
region due to the pion instability. As follows from (\ref{eq:Ijk13}), 
its contribution to the oscillation phase can be neglected provided that 
\be
\tilde{\sigma}_{x\mu}\ll 
\left(\frac{2P}{\Delta m_{jk}^2}
\frac{v_\pi}{\Gamma}\right)^{1/2}
=\,\left(\frac{1}{2\pi} l_{\rm osc}\, l_{\rm decay}\right)^{1/2}. 
\label{eq:cond5}
\ee

The second exponential factor (\ref{eq:Ijk13}) leads to the suppression of the 
oscillations in the case $(\Delta m_{jk}^2/2P)\tilde{\sigma}_{x\mu} \gg 1$,  
i.e.\ it describes possible production decoherence effects. In particular, 
for the muon neutrino survival probability in the 2-flavour case we obtain, 
assuming that (\ref{eq:cond5}) holds, 
\be
P_{\mu\mu}=c^4+s^4+2 s^2 c^2\, e^{-\frac{1}{2}
\big(\frac{\Delta m^2}{2P}\big)^2\tilde{\sigma}_{x\mu}^2}\,
\cos \Big(\frac{\Delta m^2}{2P}(L-x_S)\Big)\,. 
\label{eq:P4}
\ee
Thus, in the limit 
(\ref{eq:cond4}) the decoherence parameter is $\frac{\Delta m^2}{2P}
\tilde{\sigma}_{x\mu}$.

Finally, consider the case when the muon interacts with the medium 
but there are no muon detectors and the muon position is not measured. 
Equivalently, one can imagine that the muon position is measured but 
the correspondence between the detected neutrino and the coordinate of 
the production point of the accompanying muon is not established.  
In both these cases the neutrinos are not tagged, and one 
has to integrate (\ref{eq:Ijk11}) over $x_S$. The integration of the 
squared modulus of $g_\mu$ in (\ref{eq:Ijk11}) over $x_S$ just gives the 
normalization constant of this function which does not influence the 
oscillation probabilities, and we simply recover the results obtained 
in the case when the muon is not detected.  

{}From the above discussion of the muon detection case it is clear 
what happens if the interaction of the pions in the bunch 
between themselves or with other particles which may be present in the 
neutrino source is taken into account. Assume first that this interaction 
identifies the individual pion whose decay produces a given neutrino. 
For example, the pion decay may lead to some recoil of the neighbouring 
particles which may be detected. This would localize the coordinate of the 
neutrino production point within an uncertainty of order of the inter-pionic 
distance (or, correspondingly, the distance between the pion and the other 
neighbouring particles in the source) $r_0$, and would lead to neutrino 
tagging. The production decoherence parameter in this case is 
$(\Delta m^2/2P)r_0$. If, however, the information about the interaction 
between the decaying pion and the surrounding particles is not recorded and 
not used for neutrino tagging, the oscillations occur in exactly the same 
way as if pions did not interact with each other or with other 
particles.

\section{\label{sec:exp}Implications for neutrino oscillation 
experiments}  

We shall now estimate possible effects of production decoherence for a
number of past, ongoing and forthcoming or proposed neutrino oscillation 
experiments. 

As was pointed out in section~\ref{sec:pimu}, in realistic situations the 
spatial widths of the wave packets of parent pions are much smaller than 
all the length parameters of physical interest in the problem, therefore we 
will be using here the formulas from section \ref{sec:cohsum} 
obtained for pointlike pions. It was demonstrated in that section that in the 
case $\Gamma l_p/v_\pi\gg 1$, when most of the pions decay before reaching the 
end of the decay tunnel, the production decoherence effects are governed 
by the parameter $\xi$ defined in eq.~(\ref{eq:xi}) (see also \cite{HS}), 
whereas in the opposite case $\Gamma l_p/v_\pi \ll 1$
the decoherence parameter is $\phi_p$ defined in (\ref{eq:phi}). 
In table 1 we give the parameters $\Gamma l_p/v_\pi$, $\phi_p$ and 
$\xi$, 
along with some other relevant physical characteristics, for a number of 
experiments
\cite{LSND,KARMEN,MiniBooNE,NOMAD,CCFR,CDHS,K2K,T2K,Minos,Nova,betabeams}. 
If not otherwise specified, we assume \mbox{$\Delta m^2=2$ eV$^2$} for 
all experiments. For a number of experiments (NOMAD, CCFR, CDHS) we adopt  
the values of $\Delta m^2$ that correspond to the maximal sensitivity of  
these experiments. For $\beta$-beams we consider a short-baseline setup with 
$L=130$ km, the neutrino energy in the ion rest frame $E_0=2$ MeV, ion 
lifetime \mbox{$\tau_0=1$ s} and $\gamma=100$. 

\begin{table}[ht]
\caption{Production coherence for a number of experiments 
[12-22]. 
Unless otherwise specified, $\Delta m^2=2~\mbox{eV}^2$. 
For $\beta$-beams we adopt $E_0=2$ MeV, $\tau_0=1$ s, $\gamma=100$.}

\vspace*{2.0mm}
\hspace*{-5mm}
\begin{tabular}{lccccccccc}
\hline
Experiment & $\langle E_\nu \rangle$(MeV) & $L$(m) & $l_p(m)$ & $l_{\rm 
dec}$(m) & $l_{\rm osc}$(m) & $\Gamma l_p/v_P$ & $\phi_p$ & 
$\xi$ \\
\hline
LSND      & $\sim$40  & 30 & 0   & 0   & 50    &  -   &  0  &  0 \\

KARMEN    & $\sim$40  & 17.7 & 0   & 0  & 50  & -   & 0   & 0  \\

MiniBooNE & $\sim$800 & 541  & 50 & 89 & 992 & 0.56& 
0.32 &  0.56 \\

NOMAD 
& $2.7\cdot 10^4$ & 770  & 290 & 3009 & 33480 & 0.1& 0.054 &  0.56 \\
~~(20 eV$^2$) &  &   &  &  & 3348 & 0.1& 0.54 &  5.64 \\

CCFR($10^2\,$eV$^2$) 
 & $5\!\cdot\!10^{4}$ & 891  & 352 & 5570 & 1240 &  0.06& 1.78 &   28.2 \\

CDHS & 3000 & 130  & 52 &  334 & 3720 & 0.155 & 0.088 &  0.56 \\

~~(20 eV$^2$) &  &   &  &  & 372 & 0.155& 0.878 &  5.64 \\

K2K & $1500$ & 300  & 200 & 167 & 1861 & 1.2& 0.68 &  0.56 \\

T2K & 600 & 280  & 96 & 66.4 & 744 & 1.45& 0.81 &  0.56 \\

MINOS & 3300 & 1040  & 675 & 368 & 4092 & 1.84& 1.04 & 0.56 \\

NO$\nu$A & 2000 & 1040  & 675 & 223 & 2480 & 3.03& 1.71 &  0.56 \\

$\beta$-beams & 400 & 
$1.3\!\cdot\!10^{5}$ & 2500 & 
$3\!\cdot\!10^{10}$ & 
496 & $8.3\!\cdot\!10^{-8}$ & 31.7 & $3.8\!\cdot\!10^{8}$ \\
\hline
\end{tabular}
\end{table}

A few comments are in order. Obviously, the results we have obtained for the 
case when neutrinos are produced in pion decays are also valid for decays 
of any other parent particles, including muon decay, provided that the 
particles produced alongside the neutrino are not detected. Next, we note 
that for LSND and KARMEN experiments, in which neutrinos were produced in  
decays of parent muons at rest, the sizes of the neutrino sources  
are very small compared to the baselines and oscillation lengths and  
therefore can be considered to be essentially zero. This means that the 
produced neutrino states are characterized by large energy and momentum 
uncertainties, i.e.\ for these experiments the production coherence condition 
(\ref{eq:coh1}) is therefore satisfied very well. Notice that the parameter 
$\xi$ is practically energy independent (it depends essentially only on the 
neutrino production process and $\Delta m^2$). 
This is because the decay width in the laboratory frame $\Gamma\propto 1/E$, 
and in the denominator of (\ref{eq:xi}) its energy dependence cancels 
with that of $P\simeq E$. The only remaining energy dependence is 
through the pion velocity $v_\pi$ (or in general through the velocity 
of the parent particle $v_P$), which is very weak for relativistic 
parent particles. For $\pi\to \mu\nu$\ decay in flight and $\Delta m^2\sim 
2~{\rm eV}^2$ the parameter $\xi$ is always $\sim {\cal O}(1)$. This 
does not necessarily mean that production coherence is always violated 
in this case. Indeed, as discussed in section \ref{sec:coh}, for $l_p < 
l_{\rm decay}$ production decoherence is governed by $\phi_p$ rather 
than by $\xi$.

\begin{figure}
\hbox{\hfill
\hspace*{1.4cm}
\includegraphics[width=5.6cm]{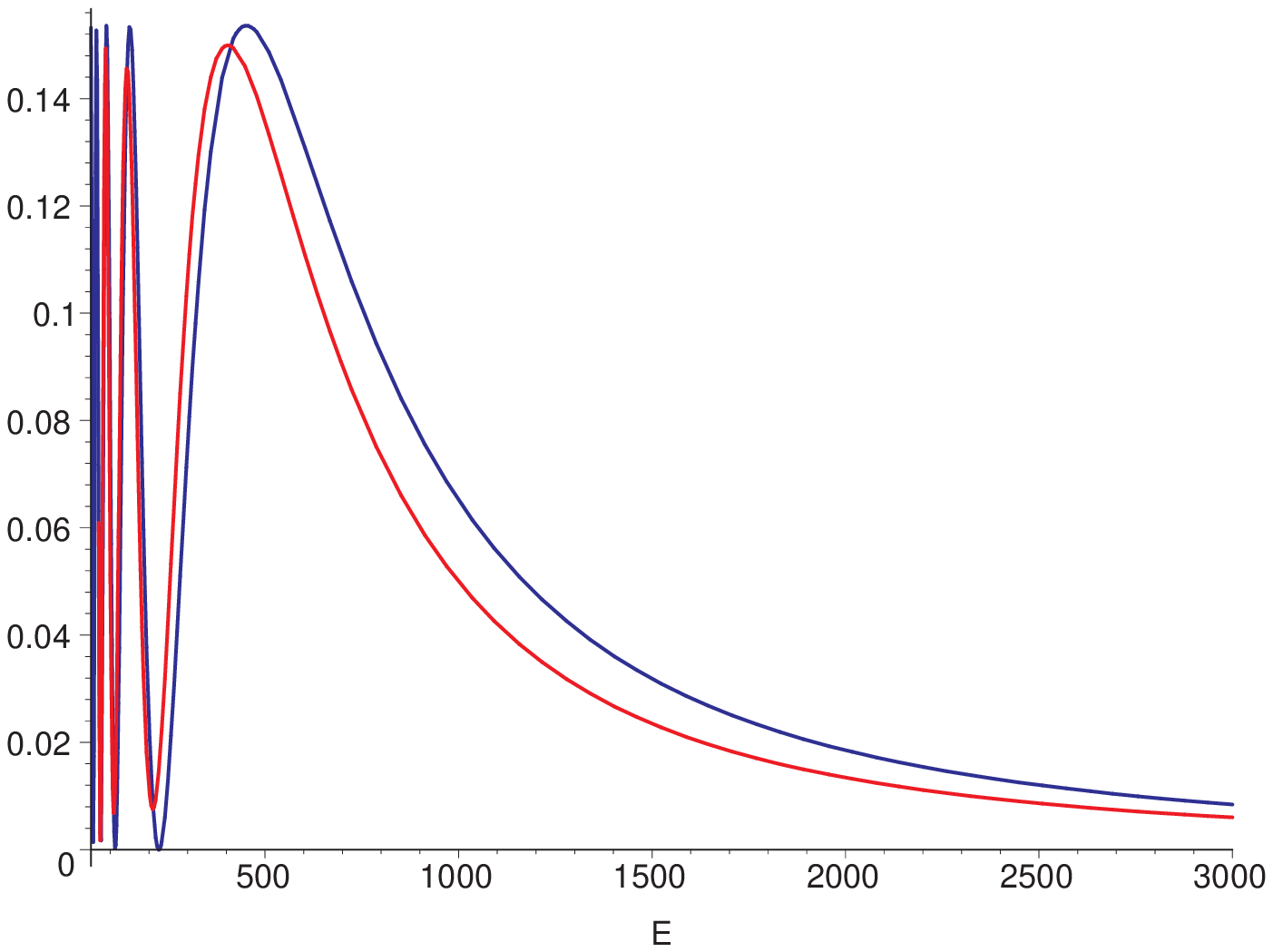}
\hspace*{0.4cm}
\includegraphics[width=5.6cm]{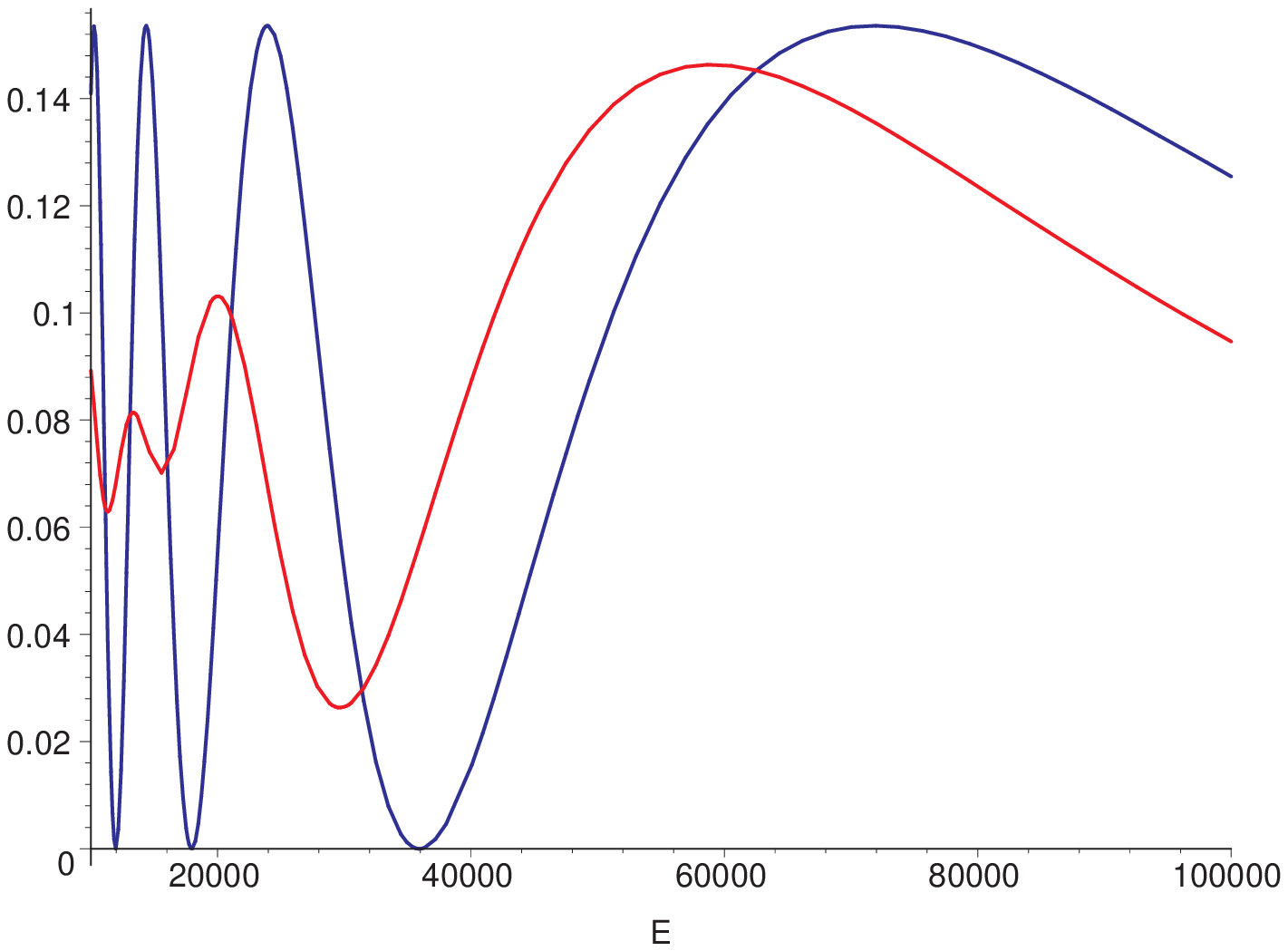}
\hfill}
\caption{Probabilities $1-P_{\mu\mu}$ for the T2K experiment (left panel) and 
the CCFR experiment (right panel) for parameters indicated in table 1 and 
$U_{\mu 4}=0.2$. Red curves: the actual oscillation probabilities, blue 
curves: the oscillation probabilities in the case of neutrino emission from 
a single point located in the middle of the decay tunnel. Neutrino 
energies are in MeV.} 
\label{fig:T2KCCFR}
\end{figure}

As can be seen from the table, production decoherence effects should be  
noticeable for MiniBooNE, NOMAD ($\Delta m^2=20$ eV$^2$), CCFR ($\Delta m^2=
100$ eV$^2$), CDHS ($\Delta m^2=20$ eV$^2$), K2K, T2K, MINOS and NO$\nu$A. 
Very large decoherence effects are expected for short-baseline $\beta$-beams 
with the parameters quoted in the table. For illustration, 
in fig.~\ref{fig:T2KCCFR} we show the oscillation probabilities $1-P_{\mu\mu}$ 
for the T2K and CCFR experiments for the parameters indicated in the table and 
$U_{\mu 4}=0.2$. The red curves show the actual oscillation probabilities, 
whereas the blue curves were obtained for neutrino emission from a 
single point located in the middle of the decay tunnel, i.e. correspond  
to the probabilities obtained neglecting the decoherence effects. 
Another example of production decoherence can be found in fig.~3 of 
ref.~\cite{Patrick}, where $\nu_e \to \nu_s$ oscillations in 
low-energy $\beta$-beam experiment with  
$\gamma=30$, $l_p=10$ m and $L=50$ m were considered. Effects of suppression 
of the oscillations due to the averaging of the neutrino production coordinate 
over the straight section of the storage ring and of the neutrino absorption 
coordinate along the detector can be clearly seen.

\section{\label{sec:disc}Discussion}

In the present paper we have studied in detail neutrino production 
coherence, which ensures that the emitted neutrino state is a coherent 
superposition of different mass eigenstates, as well as its implications 
for neutrino oscillations. We considered neutrino production using $\pi\to
\mu\nu$ decay as an example. However, our analysis also applies to neutrinos 
produced in any other decay process (such as $K$-decay, $\mu$-decay, 
$\beta$-decay, etc.), provided that particles accompanying neutrino production 
are undetected, as it is usually the case.  

We studied neutrino production coherence in two completely 
different approaches. In one of them, based on the quantum-mechanical wave 
packet formalism, we found the transition amplitude ${\cal A}_{\alpha\beta}
(L, t)$ by summing the amplitudes of neutrino production at different points 
along the path of the parent particle. The oscillation probability 
$P_{\alpha\beta}(L)$ is then found by integrating the squared modulus of the 
obtained amplitude over time. In the second approach we assumed that each 
individual neutrino production event is fully coherent, so that the 
evolution of the produced neutrino should be described by the standard 
oscillation probability. However, the neutrino emission can occur (with some 
probability) at any point within the neutrino source, and therefore the 
effective oscillation probability is obtained by the proper averaging of 
the standard probability over the neutrino source. In both approaches 
we assumed that the neutrino detection process is perfectly localized. 

We have found that in the case when the parent particles, in decays of 
which the neutrinos are produced, can be considered as pointlike, the 
results of the two approaches exactly coincide. This happens despite 
the fact that the first approach is fully quantum mechanical, while the 
second one, based on the probability summation, is essentially classical. 
We thus confirm the conclusion of ref.~\cite{HS} where, along with the 
probability summation, a simplified treatment of the quantum mechanical 
coherent amplitude summation was given. 

It should be stressed that the equivalence of the two approaches studied 
in this paper is different from the one discussed in \cite{Rich,KNW}. 
In those papers the authors have pointed out that it is impossible to 
distinguish experimentally between an ensemble of neutrinos described by wave 
packets, each with the energy distribution amplitude $g(E)$, and a beam 
of plane-wave neutrinos, each with well defined energy, but with the 
energy spectrum of the beam given by the squared modulus of the same 
function $g(E)$. In our case we establish the equivalence between a beam  
of neutrinos described by wave packets and a beam of pointlike neutrinos, 
which are just the opposites of the plane waves.

How can one understand the equivalence of the two approaches studied in 
this paper which takes place in the case of pointlike pions? Although the 
position of the pion decay (and neutrino production) point is not {\em a 
priori} exactly known, having uncertainty of order $l_p$ for $\Gamma l_p/v_\pi 
\ll 1$ and $l_{\rm decay}=v_\pi/\Gamma$ for $\Gamma l_p/v_\pi \gg 1$, the 
{\sl spatial size} of the production region for each individual production 
event is very small: it is given by the size of the smallest wave packet of 
the particles participating in neutrino production, in our case of the pion. 
For pointlike pions the space-time localization of the detection process 
actually allows one, for each detection event, to pinpoint the coordinate of 
the neutrino emission. Thus, in 
this case there is no quantum mechanical uncertainty in the coordinate 
of neutrino production and therefore no interference between 
the amplitudes of neutrino emission from different (even closely located) 
points. This explains why our results obtained through the coherent summation 
of the amplitudes of neutrino production along the neutrino source coincide 
with those found by the simple incoherent summation of the oscillation 
probabilities. 

{}From the above argument it follows that one can expect some deviations 
between the results of the coherent amplitude summation and 
incoherent probability summation approaches if pions are described by 
wave packets of finite size $\sigma_{x\pi}$. This is indeed confirmed by our 
treatment of the finite $\sigma_{x\pi}$ case in sections 
\ref{sec:finlength} and \ref{sec:altern}. We have found that for small 
$A_\pi\sim [v_g/(v_g-v_\pi)](\Delta m^2/2P)\sigma_{x\pi}$
the oscillation probabilities get extra oscillatory terms proportional to 
$A_\pi$. For different shapes of the pion wave packets these corrections 
differ only by 
numerical factors (cf. eqs.~(\ref{eq:ApiGauss}) and ~(\ref{eq:ApiBox})). 
At the same time, for $A_\pi \gg 1$ the neutrino production coherence is 
violated, and the oscillatory terms in the oscillation probabilities are 
strongly suppressed. From the above discussion it should be rather obvious 
that the amplitude summation and probability summation approaches should lead 
to different results also in the case when the detection process is not 
perfectly localized, i.e.\ when the particles participating in neutrino 
detection are described by wave packets of finite size. This should be the 
case even if the production process is perfectly localized. 

\begin{figure}
\begin{center}
\includegraphics[width=8.0cm]{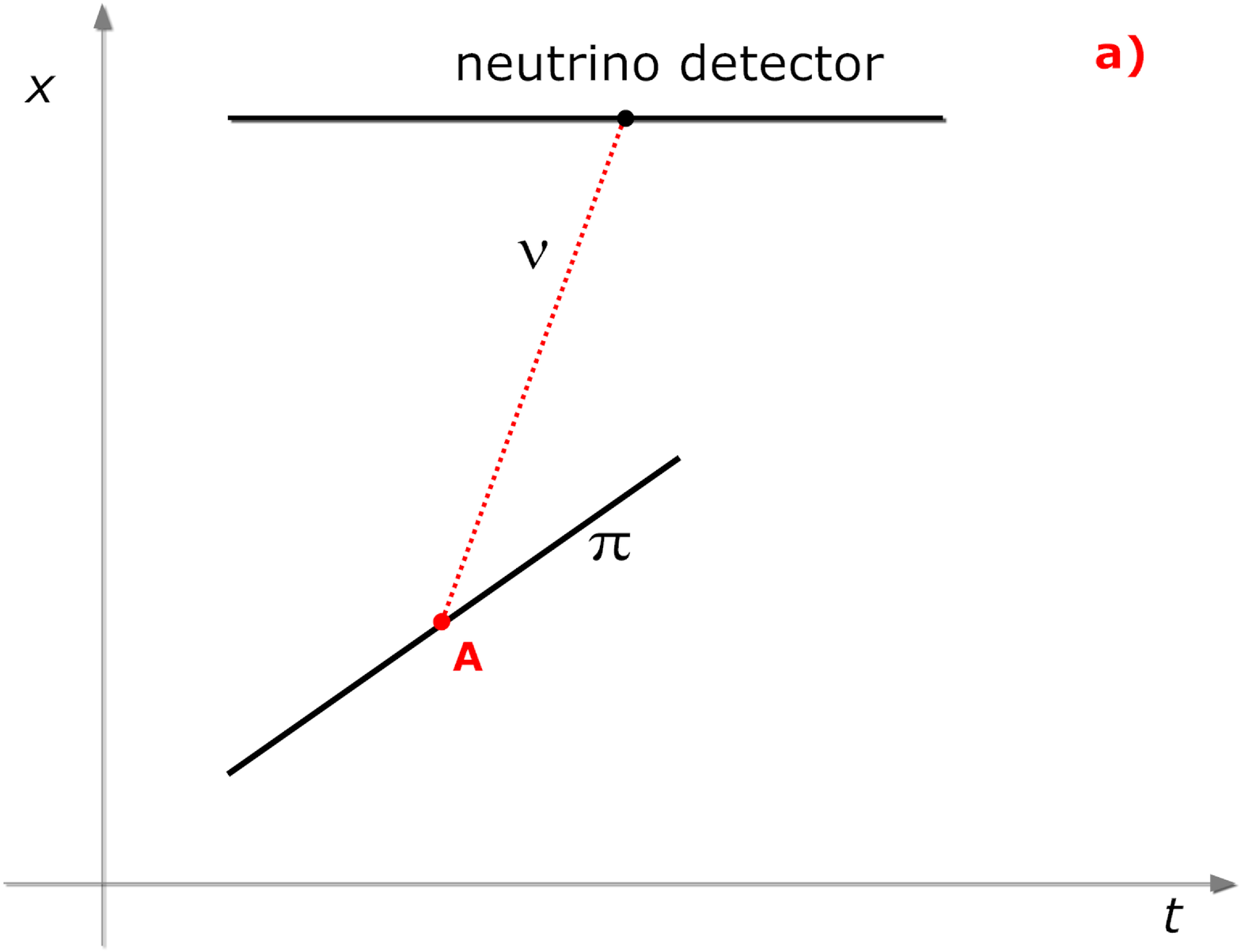}
\includegraphics[width=8.0cm]{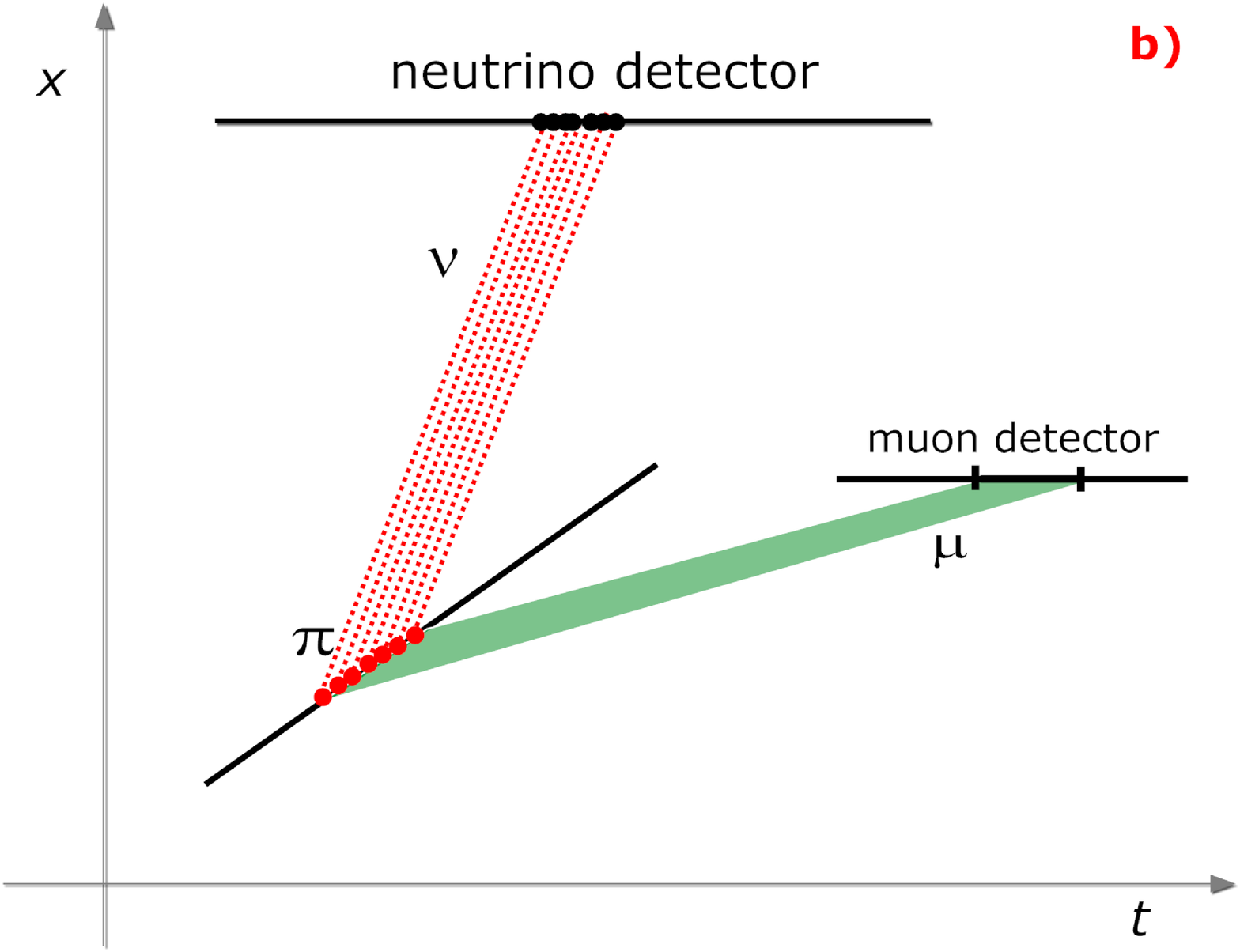} \\

\includegraphics[width=8.0cm]{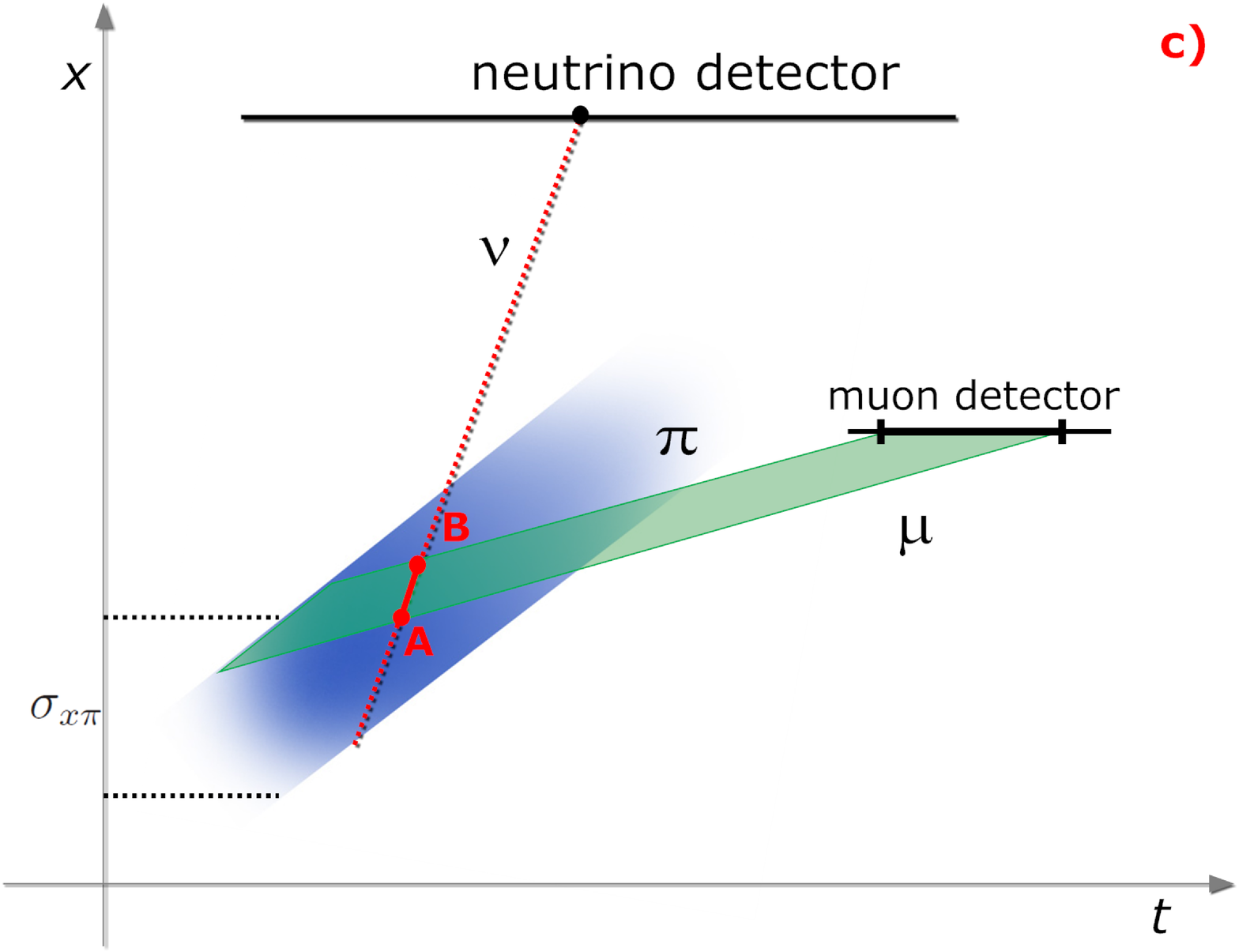} 
\includegraphics[width=8.0cm]{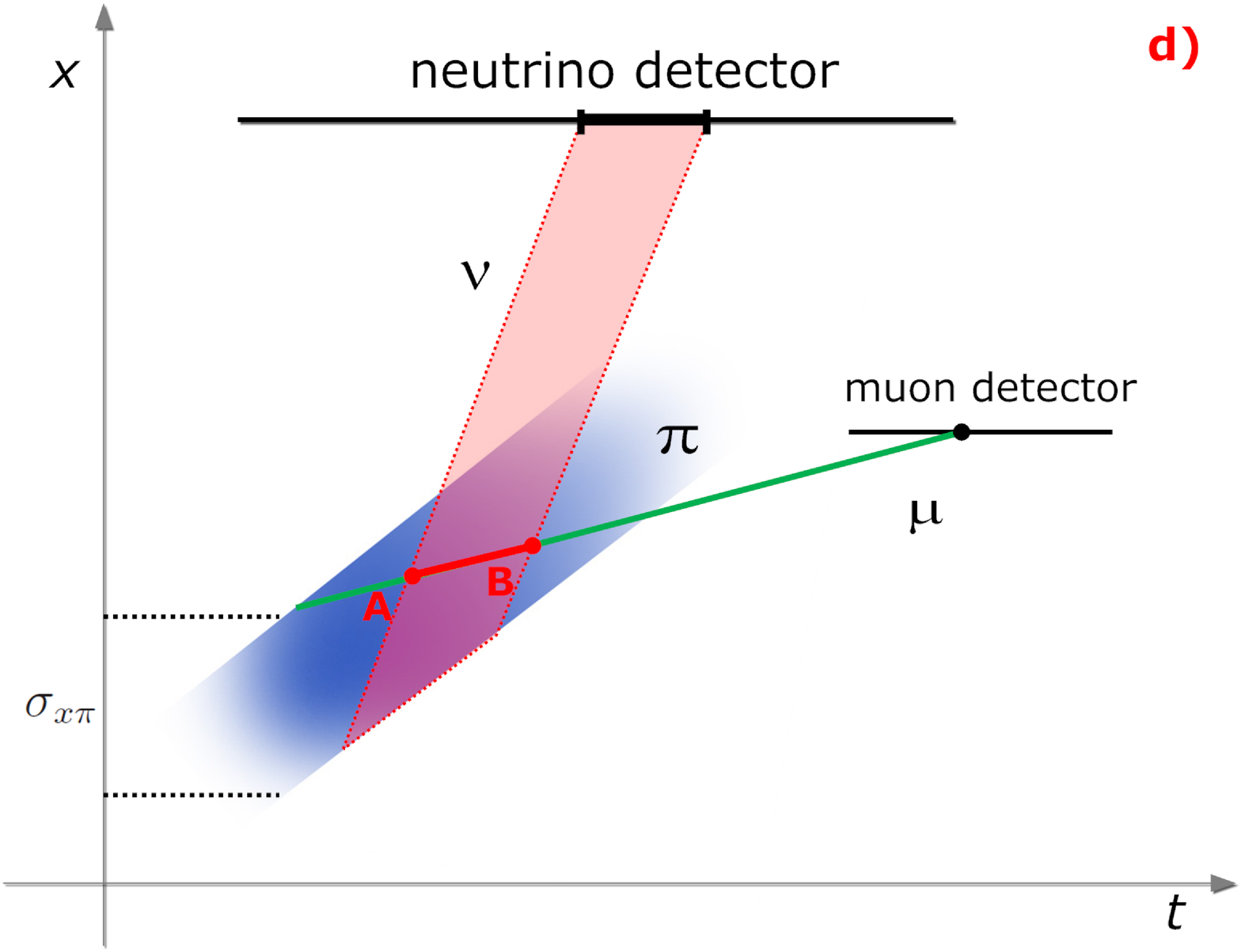} \\

\includegraphics[width=8.0cm]{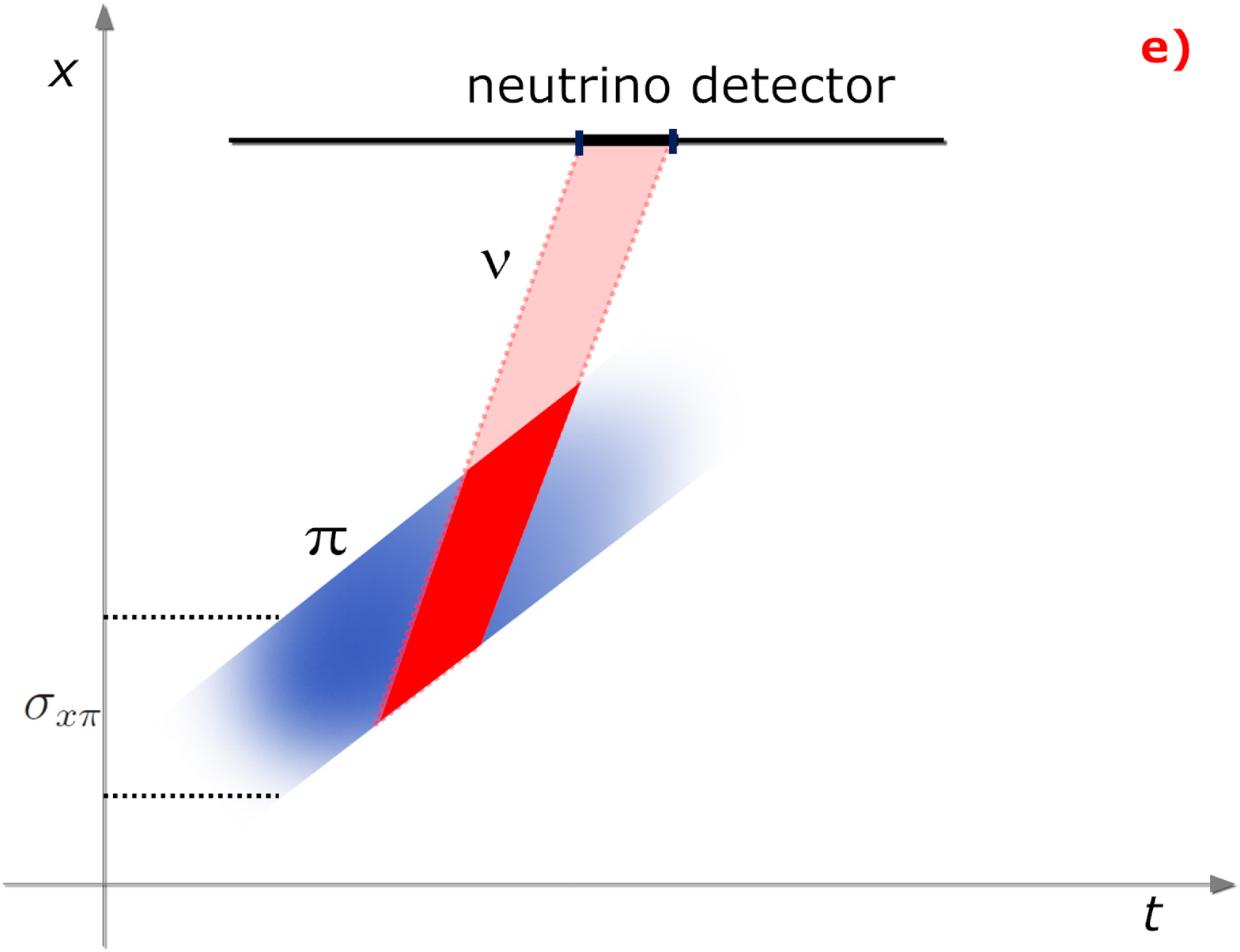} 
\includegraphics[width=8.0cm]{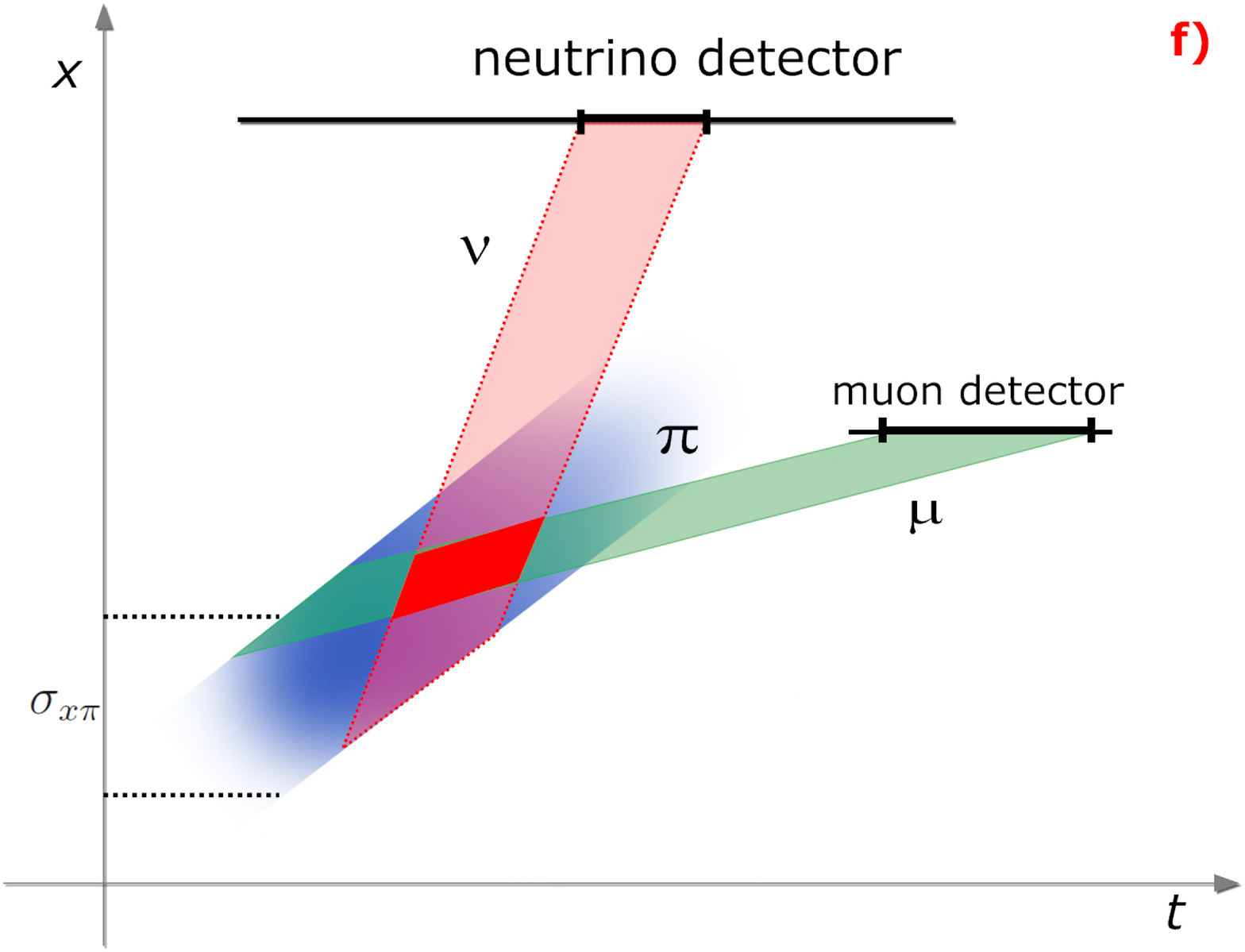}
\end{center}
\caption{Space-time diagrams corresponding to six experimental setups. The 
regions of coherent production in each case are shown in deep red (see text).} 
\label{fig:figa}
\end{figure}

The above points as well as the role of detection of the charged lepton 
produced alongside the neutrino are illustrated by fig.~\ref{fig:figa}. In 
this figure we present the space-time diagrams that correspond to six 
different experimental setups in the case of neutrinos produced 
in $\pi\to \mu\nu$ decays. The size of the region of coherent amplitude 
summation in each case is determined by the interplay of three factors, 
namely, whether the pion can be considered as pointlike or not and whether 
the neutrino and muon detection regions can be considered to be space-time 
points or not.%
\footnote{Clearly, this can be made more precise. For the pion, one 
needs to compare $\sigma_{x\pi}v_g/(v_g-v_\pi)$ with $l_{\rm osc}$. 
For the detection processes, it is necessary to compare the size of the  
detection region 
with $l_{\rm osc}$. }
%
Whenever this region degenerates to a point, the coherent and incoherent 
summation approaches yield identical results. For simplicity, in the 
cases when the neutrino and the muon detection processes are not fully 
localized, we display them as being extended in time but localized in 
space. More general situations where these processes are delocalized 
both in space and time can be readily studied; however, the 
corresponding results will not modify our qualitative conclusions.  
The six panels in fig.~\ref{fig:figa} thus illustrate the following 
situations: 
\begin{itemize} 
\item[a)] 
The pion and the neutrino detection region are both pointlike, the muon 
goes undetected. The first two conditions are sufficient to identify the point 
where the neutrino was produced, thus eliminating any quantum mechanical 
uncertainty in the emission coordinate. The oscillation probability 
found through the coherent amplitude summation must be identical to the one 
obtained by incoherent probability summation. 

\item[b)] 
The pion and the neutrino detection region are both pointlike, the muon 
detection region has a finite size. The latter does not restore coherence  
of neutrino emission from different points (illustrated by the dotted 
lines in the figure). The corresponding contributions have to be summed at 
the level of probabilities. 

\item[c)] The pion is of finite extension and so is the muon 
detection region, whereas the neutrino detection is pointlike. A 
one-dimensional region $AB$ is formed as the intersection of the 
regions corresponding to these three conditions. It is impossible in principle 
to determine from which point in the segment $AB$ the neutrino was emitted, 
and the amplitudes of neutrino emission from all such points thus interfere. 
The amplitude summation produces the results that are different from those 
found through the probability summation. 
 
Depending on the degree of its delocalization, the muon detection process may 
reduce the length of the segment $AB$ compared to the case when the muon is 
undetected (cf.~figs.~\ref{fig:scheme} and \ref{fig:figa}c) 
and thus diminish the effects of averaging of neutrino oscillations 
caused by neutrino production decoherence.

\item[d)] The pion is of finite extension, the neutrino detection region is 
also finite, but the muon detection region is pointlike. As in c), the region 
of amplitude summation is a segment, and there is no equivalence with  
incoherent probability summation.

\item[e)] Both pion and neutrino detection regions are of finite 
size, while the muon goes undetected. A 2-dimensional region 
of amplitude summation is formed; the amplitudes of neutrino emission 
from different points of this region interfere.  

\item[f)] Same as in e), but with muon detection (finite detection region).
As in case e), the region of amplitude summation is 2-dimensional.  
Its size may be smaller than in case e) due to constraints from 
the muon detection. 

\end{itemize}
Not shown in fig.~\ref{fig:figa} is one more case when the 
results of the amplitude summation and probability summation 
approaches  coincide: neutrino detection region is of finite 
extension, whereas the pion is pointlike and the muon detection is 
perfectly localized in space an time. 

Are deviations between the results of the coherent amplitude summation 
and incoherent probability summation approaches experimentally observable? 
Consider the case of perfectly localized neutrino detection, non-zero spatial 
widths of the pion wave packets and no muon detection 
(fig.~\ref{fig:scheme}). The corrections to the oscillation probabilities 
due to $\sigma_{x\pi}\ne 0$ are governed in this case by the parameter 
$A_\pi$ which, as follows from~(\ref{eq:ApiGauss}) or (\ref{eq:ApiBox}), can 
become sizeable for extremely high energies of the parent pion. This requires 
\be
2 \left(E_\pi \sigma_{x\pi}\right)
\frac{\Delta m^2}{m_\pi^2} \gtrsim 1\,.
\label{eq:condit}
\ee
For instance, if the width of the proton wave packets $\sigma_{xp}\sim 
10^{-4}$ cm, one can expect also $\sigma_{x\pi}\lesssim 10^{-4}$ cm 
(see eq.~(\ref{eq:sigmaxpi})); for $\Delta m^2 \sim 1$ eV$^2$ the parameter 
$A_\pi$ will then become of order one for pion energies $E_\pi\gtrsim 
10^3$ TeV. Such energies are not feasible, and even if they were, the 
corresponding neutrino oscillation lengths would be far too large for 
any oscillation experiment. Another possibility to make $A_\pi$ sizeable 
would be to increase significantly the spatial width of the wave packets of 
ancestor protons, which would in turn increase the values of $\sigma_{x\pi}$. 
However, it is not clear how this can be achieved.%
\footnote{It might still be possible, however, to achieve relatively 
large widths of the wave packets of the parent particles if neutrinos 
are produced in muon decays.}    
 
The two approaches to production decoherence effects that we followed in 
this paper describe the same physical phenomenon -- the suppression of the 
oscillating terms in neutrino transition and survival probabilities 
due to delocalization of the production process. The nature of this 
suppression, however, is different in different situations. 

In general, the proper description of the decoherence phenomenon requires 
using the quantum-mechanical approach with coherent summation of the 
amplitudes of neutrino production at different points. In this framework one 
considers the violation of production coherence due to lack of localization 
of each individual neutrino production event. This delocalization of neutrino 
production is related to the fact that the exact coordinate of the neutrino 
emission point is not known. The decoherence parameter is therefore 
essentially the ratio of the size of the neutrino production region and 
the neutrino oscillation length. As was shown in section \ref{sec:cohsum}, in 
the limit $\Gamma l_p/v_\pi \ll 1$ the decoherence parameter is $\phi_p
=2\pi l_p/l_{\rm osc}$, whereas for $\Gamma l_p/v_\pi \gg 1$
it is $\xi=2\pi l_{\rm decay}/l_{\rm osc}$. 
The decoherence parameters can also be represented as the ratio of the energy 
difference of different neutrino mass eigenstates and the energy uncertainty 
inherent in the neutrino production (see section~\ref{sec:coh}): 
$\Delta E/\sigma_E\approx \phi_p$ for $\Gamma l_p/v_\pi \ll 1$; 
$\Delta E/\sigma_E\approx \xi$ for $\Gamma l_p/v_\pi \gg 1$.  
For $\Delta E/\sigma_E\gtrsim 1$ the production process can discriminate 
between different neutrino eigenstates, leading to a loss of their 
coherence. 

It is interesting to note that for decay of pointlike parent 
particles at rest ($v_P=0$) the production decoherence parameters vanish: 
the parameter $\xi$ vanishes because it is proportional to $v_P$, 
whereas $\phi_p$ is essentially zero because the effective size of the 
neutrino source $l_p$ is negligibly small for decay at rest. 
Physically, the reason for a perfect production coherence in the case
$v_\pi \to 0$ (and good spatial localization of the neutrino 
detection process) is that the oscillation baseline is fixed and, in 
particular, is independent of the time when the parent pions decay.
The oscillation baseline is not, however, fixed if the wave packets of 
the parent particles are of finite size; in this case the averaging of 
the oscillation phase over the widths of these wave packets can lead to 
decoherence effects.

In the second approach considered in the present paper it is assumed 
that the position of each neutrino emission point is exactly known. 
Each individual 
neutrino production event is assumed to be fully coherent, and the 
averaging of the production coordinate over the neutrino source is 
performed simply because the source is extended and the neutrino 
emission can occur in any place inside it. The averaging in this case is 
done at the probability level and is described by the parameter $\phi_p$ 
for $\Gamma l_p/v_\pi \ll 1$ and by $\xi$ for $\Gamma l_p/v_\pi \gg 1$. 
These parameters are just related to the effective size of the neutrino 
source, which in these limits is, respectively, $l_p$ and $l_{\rm decay}=
v_\pi/\Gamma$. Actually, there is no notion of production decoherence in this 
case, and the suppression of the oscillatory terms in the probabilities is 
merely due to the averaging of the standard oscillation probability over the 
macroscopic sizes of the neutrino source and detector. 

As we have discussed in detail, the two approaches  
turn out to be equivalent if the detection process is perfectly  localized 
in space and time and in addition the spatial size of the wave packet of the 
parent particle can be neglected. This conclusion does not change if the 
particles accompanying the neutrino production (such as muon in $\pi\to 
\mu\nu$ decay) interact with the medium or are directly detected. 
However, if the  detection of the accompanying particles is 
used for neutrino tagging, i.e.\ allows one to establish the coordinate 
of the neutrino emission point in the source with an accuracy 
$\sigma_{x\mu}$, this can affect the oscillation probabilities by 
reducing the averaging effects. In the case when the coherent amplitude 
summation does not reduce to the probability summation (e.g., for 
finite-size pion wave packets), this happens due to a better localization of 
individual neutrino production points (assuming that $\sigma_{x\mu} 
< l_p$, $l_{\rm decay}$), which improves the production coherence.
In the case when the use of the probability summation is legitimate, 
neutrino tagging through the detection of the accompanying particles would 
just mean that the neutrinos are emitted not from the whole neutrino source, 
but from a region of it of the effective size $\sigma_{x\mu}$. The integration 
over the coordinate of the neutrino production should then be carried 
out only over this region. 

In conclusion, we have identified  the condition under which coherent and 
incoherent summations over points in the neutrino production region yield 
different oscillation results. This condition can be succinctly stated as 
follows: the two approaches lead to different results whenever the 
localization properties of the parent particles at neutrino production 
and of the detection process 
are such that they prevent the precise localization of the 
point of neutrino emission. The difference in the  oscillation results is 
negligible for present accelerator experiments, and therefore the standard 
averaging of the oscillation probabilities over the finite spatial extensions 
of the neutrino source and detector properly takes decoherence effects into 
account. Whether it is possible to devise a realistic experiment that 
could probe this difference remains to be seen. 
	
The authors are grateful to Thomas Schwetz and Manfred Lindner for 
useful discussions.

\end{document}